\documentclass[journal]{IEEEtran}

\usepackage{amsmath,amsfonts,bm}

\def\eqref#1{equation~\ref{#1}}

\def\floor#1{\lfloor #1 \rfloor}
\def\1{\bm{1}}

\DeclareMathAlphabet{\mathsfit}{\encodingdefault}{\sfdefault}{m}{sl}
\SetMathAlphabet{\mathsfit}{bold}{\encodingdefault}{\sfdefault}{bx}{n}

\usepackage{cite}
\usepackage{graphicx}
\usepackage{subfigure}
\usepackage{url}
\usepackage{xcolor} 
\usepackage{caption} 
\usepackage{multirow}
\usepackage{array} 

\begin{document}

\title{CLEEGN: A Convolutional Neural Network for Plug-and-Play Automatic EEG Reconstruction}

\author{
Pin-Hua Lai, Bo-Shan Wang, Wei-Chun Yang, Hsiang-Chieh Tsou, and Chun-Shu Wei,~\IEEEmembership{Member,~IEEE}
\thanks{This work was supported in part by the National Science and Technology Council (NSTC) under Contracts 109-2222-E-009-006-MY3, 110-2221-E-A49-130-MY2, 110-2314-B-037-061, 112-2321-B-A49-012, and 112-2222-E-A49-008-MY2; and in part by the Higher Education Sprout Project of National Yang Ming Chiao Tung University and Ministry of Education. Corresponding author: Chun-Shu Wei (wei@nycu.edu.tw).}
\thanks{
Pin-Hua Lai, Wei-Chun Yang, Hsiang-Chieh Tsou, and Chun-Shu Wei are with Department of Computer Science, National Yang Ming Chiao Tung University (NYCU), Hsinchu, Taiwan. Chun-Shu Wei is also with Institute of Biomedical Engineering and Institute of Education, NYCU, Hsinchu, Taiwan.}
\thanks{
Bo-Shan Wang is with Institute of Biomedical Engineering and Institute of Education, NYCU, Hsinchu, Taiwan.}
}

\markboth{Journal of \LaTeX\ Class Files,~Vol.~14, No.~8, January~2024}%
{Shell \MakeLowercase{\textit{et al.}}: Bare Demo of IEEEtran.cls for IEEE Journals}

\maketitle

\begin{abstract}
Human electroencephalography (EEG) is a brain monitoring modality that senses cortical neuroelectrophysiological activity at high temporal resolution. One of the greatest challenges posed in applications of EEG is the unstable signal quality that is susceptible to inevitable artifacts during recordings. To date, most existing techniques for EEG artifact removal and reconstruction are applicable to offline analysis solely, or require individualized training data to facilitate online reconstruction. We have proposed CLEEGN, a novel light-weighted convolutional neural network for plug-and-play automatic EEG reconstruction. CLEEGN is based on a subject-independent pre-trained model using existing data and can operate on a new user without any further calibration. The performance of CLEEGN was validated using multiple evaluations including waveform observation, reconstruction error assessment, and decoding accuracy on well-studied labeled datasets. The results of simulated online validation suggest that, even without any calibration, CLEEGN can largely preserve inherent brain activity and outperforms leading artifact removal methods in the decoding accuracy of reconstructed EEG data. In addition, visualization of latent features reveals the model's behavior and the intriguing process of artifact removal in EEG data. We foresee pervasive applications of CLEEGN in prospective works of online plug-and-play EEG decoding and analysis.
\end{abstract}

\begin{IEEEkeywords}
EEG, Brain-computer interface, EEG artifact removal.
\end{IEEEkeywords}

\section{Introduction}

\IEEEPARstart{S}{ince} the first record of human electroencephalogram (EEG) performed almost a century ago (in 1924), EEG has been one of the most widely used non-invasive neural modalities that monitors brain activity in high temporal resolution \cite{koike2013near,mehta2013neuroergonomics,sejnowski2014putting}. Among a variety of modalities, EEG has extensive use in the clinical assessment of neurological and psychiatric conditions, as well as in the research of neuroscience, cognitive science, psychology, and brain-computer interfacing (BCI) \cite{nicolas2012brain}. 

EEG signals measure subtle fluctuations of the electrical field driven by local neuroelectrophysiological activity of a population of neurons in the brain cortex \cite{cohen2017does}. While the electrodes are placed on the surface of the scalp, undesired artifacts may introduce interruption in the measurements and distort the signal of interest. Even in a well-controlled laboratory with a well-trained subject who is able to maximally keep the body still and relaxed, the EEG signals, unfortunately, could be contaminated by inevitable behavioral and physiological artifacts such as eye blinks, reflective muscle movements, ocular activity, cardiac activity, etc \cite{Sazgar2019,croft2000removal,wallstrom2004automatic,romero2008comparative}. In practice, it is difficult to identify and track the sources of artifacts entirely due to their diversity and non-stationarity. Noise cancellation and artifact removal remain major issues in EEG signal processing and decoding.

Currently, numerous methods have been proposed to alleviate the influence of artifacts in EEG signals \cite{Cmp_of_Diff_Lin_Filt_4OA,AdptFilt_Auto_Corr_of_EOG,Eva_of_EMD_4EEG,Env_Interp_in_EMD,Rm_OA_Use_DWT,CCA_2_Rm_MA,ICA_of_EEG,Eva_ASR_4_AR_EEG,nazareth2006wica,zeng2016eemdica,chen2019eemdcca}. According to previous meta-analyses on EEG artifact removal literature \cite{uriguen2015eeg,jiang2019removal}, independent component analysis (ICA) is especially popular. It is majorly used in $45\%$ of EEG denoising literature. ICA-based artifact removal estimates the component activity by unmixing the EEG data in the channel domain. Through manual or automatic identification, one can exclude the artifact components and then reconstruct the EEG data through back projection based on non-artifact components.

With the fast growth of deep learning (DL) techniques \cite{lecun2015deep}, many studies have proposed innovate neural-network structure for EEG analysis or decoding \cite{pan2022matt,wei2019spatial,lawhern2018eegnet,shallow2017}. On top of that, recently developed NN-based EEG artifact removal schemes have drawn much attention as they facilitate 1) automatic artifact removal without manual interventions and 2) enhanced performance compared to conventional schemes. However, the majority of existing deep learning-based EEG reconstruction methods often lack a thorough consideration of real-world practicality. Specifically, these current methods tend to validate their effectiveness on synthetic data generated through a hand-crafted synthesis of pre-defined noiseless EEG data and noise/artifact components \cite{leite2018deep,SUN2020108,EEGdenoiseNet}, but they often fall short in providing a thorough evaluation of their capability to enhance the quality of real EEG signals. Furthermore, a practical EEG reconstruction technique intended for real-world use must also exhibit computational efficiency to support real-time signal processing pipelines. While there have been some studies that have made progress in either developing lightweight network architectures or conducting validation on real EEG data \cite{lee2020eeg,ieee9605576,CHUANG2022119586}, no prior work has successfully met the dual criteria of real-world EEG reconstruction, encompassing both validity and computational efficiency.

\begin{figure*}[ht]
    \centering
    \includegraphics[width=1.0\linewidth]{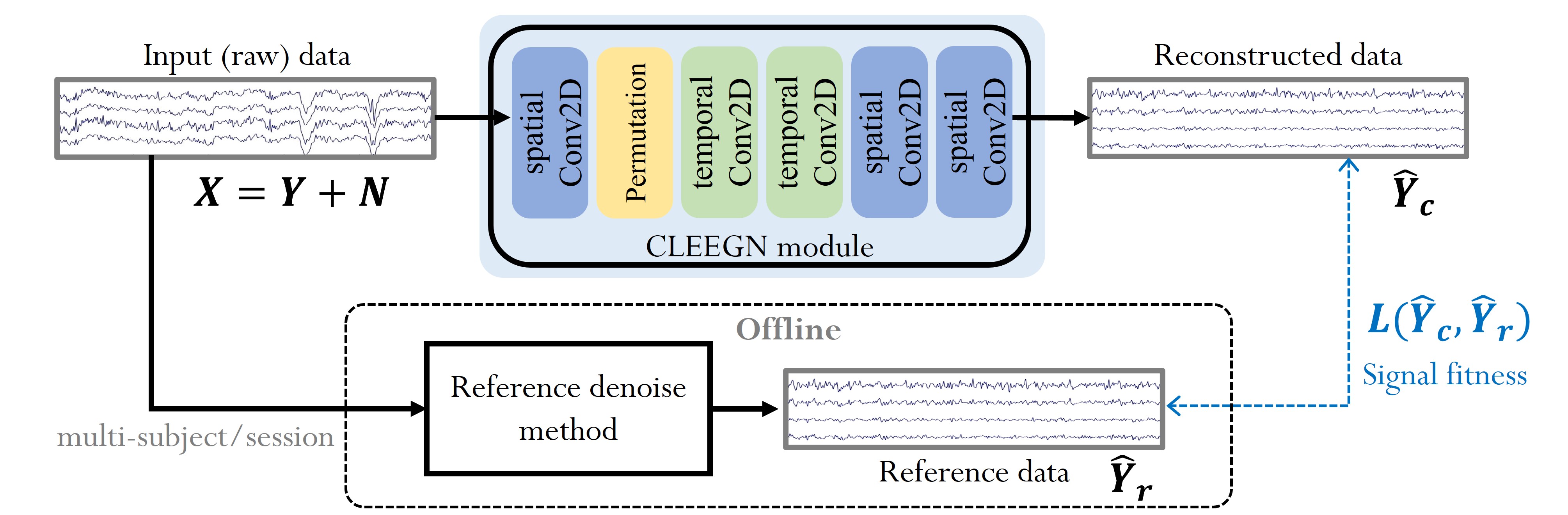}
    \caption{The denoising flow diagram with the proposed artifact removal architecture. $X$, $Y$, $N$ denote the observed EEG data, the ideal EEG signals and the noise respectively. The reference EEG data, $\tilde{Y}_r$, is the noiseless estimation generated by existing artifact removal method in the offline scenario.}
    \label{fig:cleegn_n_objective}
\end{figure*}

In this work, we propose CLEEGN, a ConvoLutional neural network for EEG reconstructioN. CLEEGN is capable of subject-independent EEG construction without any training/calibration for a new subject. The contributions of this work are three-fold:
\begin{itemize}
    \item We present CLEEGN, a lightweight convolutional neural network capable of fully automated end-to-end EEG reconstruction.
    \item CLEEGN demonstrates superior performance compared to leading online and offline methods in terms  of enhancing EEG signal quality, as evidenced by improved decoding accuracy.
    \item We introduce a framework that leverages offline reference denoising methods to enable subject-independent training, achieving zero-calibration and plug-and-play EEG reconstruction.
\end{itemize}

\section{Related Works}

Current processing techniques for EEG artifact removal are diverse based on different assumptions about the EEG signals characteristics. Depending on the assumptions and the algorithms, the developed techniques can be roughly categorized as filter-based, domain-based and neural-network based method.

Filter-based methods have been widely employed for their low computational cost and interpretability. Earlier attempts of EEG denoising employed the digital filter to eliminate the undesired signals in some certain frequency bands \cite{Cmp_of_Diff_Lin_Filt_4OA}. Another approach, adaptive filtering \cite{AdptFilt_Auto_Corr_of_EOG}, estimates artifact signals through additional EOG, EMG, ECG channels and removes these noisy signals from the recording signals by regression. Despite the efficiency, the digital filter is unable to cope with the overlapping spectrum between the EEG and the artifacts \cite{TF_Anz_of_EOG}, and the requirement for additional auxiliary electrodes in the adaptive filtering causes inconvenience.

Instead of separating EEG signals and artifacts based on the disparity in either time or frequency domain, domain-based methods separate artifacts by transforming the signals to a component domain, such as empirical mode decomposition (EMD) \cite{Eva_of_EMD_4EEG,Env_Interp_in_EMD}, wavelet transform (WT) \cite{Rm_OA_Use_DWT}, canonical correlation analysis (CCA) \cite{CCA_2_Rm_MA}, independent component analysis (ICA) \cite{ICA_of_EEG}, artifact subspace reconstruction (ASR) \cite{Eva_ASR_4_AR_EEG} and so on.

The domain-based methods can separate the artifact components better and have become more prevalent recently compared to the filter-based methods for EEG signals processing, yet the performance of these methods are heavily influenced by the hyperparameters and criteria provided by the users. For instance, the wavelet method relies on the selection of the mother wavelet function, the EMD-based method hinges on the extrema points searching algorithm and the stopping criterion, the CCA method relies on the chosen lowest autocorrelated components, and the ASR method relies on the cutoff parameter. Another factor affecting the performance of domain-based methods is manual inspection, which determines whether a independent components (ICs) in ICA and intrinsic mode functions (IMFs) in EMD methods is an artifact or brain activity. Additionally, traditional artifact removal methods are often limited to offline scenarios or struggle to deliver satisfactory performance in online contexts. Hybrid methods developed later, such as w-ICA, EEMD-ICA, EEMD-CCA, and others \cite{nazareth2006wica,zeng2016eemdica,chen2019eemdcca}, demonstrate improved performance by leveraging advantages from multiple techniques. However, these methods necessitate more complex manual parameter tuning.

Recently, neural network-based methods have been proposed to remove artifacts for EEG data. A variety of network structures have been applied to the framework for removing EEG artifacts and reconstructing clean EEG. A deep convolutional autoencoder \cite{leite2018deep} can enhance the peak-signal-to-noise ratio compared to the linear filtering method via a common CNN autoencoder structure, which has been widely used on image denoising. Their work shows that it seems practicable to transform the EEG waveform through a CNN structure. Later on, a combined framework integrating Bayesian deep learning and ICA \cite{lee2020eeg} used thresholding of the EEG data distribution to discard ICs classified as ocular artifacts. These methods leverage the flexibility of deep learning model design and achieve improvements in their assessments. Considering the non-stationary property in EEG data and the degradation phenomenon during training, 1D-ResCNN \cite{SUN2020108} was proposed, which adopted an inception-residual module in the network structure. This network is able to remove EOG, ECG, EMG on single-channel synthesis EEG data. IC-U-Net \cite{CHUANG2022119586}, unlike above-mentioned neural network methods that are trained on synthetic data, learns the reconstruction from pairs of noisy and noiseless EEG samples prepared by offline ICA approaches. The IC-U-Net can reconstruct EEG signals with enhanced SNR and increased number of independent components classified as 'brain components' on the ICLabel \cite{pion2019iclabel} method. This work showed that their reconstructed signal has higher SNR and can surely increase the number of brain components classified by ICLabel.

\section{Materials and Methods}

\subsection{Model Architecture}

CLEEGN is an encoder-decoder convolutional neural network designed to map multi-channel noisy EEG into a latent space and reconstruct it into noiseless EEG signals. The conceptual diagram with the architecture of our proposed model is presented in Figure \ref{fig:cleegn_n_objective}.

The encoder in CLEEGN structure is inspired by an existing EEG recognition CNN model \cite{wei2019spatial} which incorporates convolution blocks to capture spatiotemporal characteristics from EEG data efficiently. Considering a scenario where we possess a multi-channel EEG data acquired using an EEG headset with $C$ electrodes. The first convolution block is used to extract spatial EEG features through a convolutional layer containing $C$ spatial filters whose shape is $(C, 1)$. Following that, a permutation layer is employed to switch first and second dimension of the EEG spatial components for the subsequent block. The final block in the encoder primarily concentrates on extracting temporal information. It consists of a convolutional layer with $N_F$ temporal filters with shape $(1, \floor{f_s \times 0.1})$, where $f_s$ denotes the sample frequency of the EEG signals.

\begin{table}[h]
\caption{The architecture of CLEEGN.}
\label{tb:cleegn_architecture}
\centering
\begin{tabular}{lllll}
    \hline
    Block & Layer & \#kernels & Size & Output shape \\
    \hline
        Encoder & Input &  &  & ($B$, 1, $C$, $T$)\\
          & Conv2D & $C$ & ($C$, 1) & ($B$, $C$, 1, $T$)\\
          & Permute &  &  & ($B$, 1, $C$, $T$)\\
          & BatchNorm &  &  &\\
          & Conv2D & $N_F$ & (1, $\floor{f_s/10}$) & ($B$, $N_F$, $C$, $T$)\\
          & BatchNorm &  &  &\\
        Decoder & Conv2D & $N_F$ & (1, $\floor{f_s/10}$) & ($B$, $N_F$, $C$, $T$)\\
          & BatchNorm &  &  &\\
          & Conv2D & $C$ & ($C$, 1) & ($B$, $C$, $C$, $T$)\\
          & BatchNorm &  &  &\\
          & Conv2D & 1 & ($C$, 1) & ($B$, 1, $C$, $T$)\\ 
    \hline
    \multicolumn{5}{l}{\textsuperscript{}{$C$: \# channels, $T$: \# time points, $f_s$: Sampling rate, $B$: Batch size}}
\end{tabular}
\end{table}

As for the decoder part, we design an approximately symmetric structure to the encoder with three convolution blocks. The first convolution block decodes the EEG feature by utilizing a convolutional layer that employs $N_F$ temporal kernels, each having a shape of $(1, \floor{f_s \times 0.1})$. The subsequent second block then focuses on reconstructing the spatial information from the latent space through a convolutional layer with $C$ spatial kernels of shape $(C, 1)$. The final convolution block serves the purpose of mapping the feature domain back to the original time domain. To preserve the correspondence between the model's inputs and outputs, each convolutional layer, except for the first one in the encoder, applies zero-padding. Besides, we employed batch normalization on the components from each convolutional layer to enhance gradient stability and optimize the network more efficiently \cite{batchnorm}. The model parameters and the shape of layer output are further detailed in Table \ref{tb:cleegn_architecture}. 

\subsection{Choice of the Reference Denoise Method}

The objective of designing an artifact removal method is to reconstruct the noisy EEG data to noiseless EEG data. Currently, the generation of noisy-noiseless pair of existing neural network-based methods can divide into two categories. One approach involves synthesizing using clean EEG signals and physiological noise \cite{leite2018deep,EEGdenoiseNet,SUN2020108}. This method can generate a large scale training data faster and make up the majority in artifact removal method; however, it also has some potential issues. First, there is currently no method that can perfectly annotate clean EEG segments. Second, there is currently no research indicating the interaction between different physiological noise and brain activity. These reasons might make the developed method challenging to directly apply to real-world EEG data.

The second method for generating noisy-noiseless pair is to process the recorded EEG signals using traditional denoising techniques \cite{lee2020eeg,ieee9605576,CHUANG2022119586}. We utilize the second method which is based on simulating actual denoising behaviors, making it more closely aligned with real-world application scenarios compared to the first method. In this work, the employed traditional denoising techniques are independent component analysis (ICA) \cite{ICA_of_EEG} and artifact subspace reconstruction (ASR) \cite{Eva_ASR_4_AR_EEG}.

ICA is one of the most well-known blind source separation (BSS) method \cite{BSS_Review} with the development assumed that the recording EEG signals are linear combined from noise sources and the brain neurons. Recently developed ICLabel \cite{pion2019iclabel} can label the ICs provenance into seven different categories: brain, eye, heart, muscle, line noise, channel noise, and other. ASR is another automatic approach, which is based on the principal component analysis (PCA) method. The ASR method selects relatively noiseless periods from the multi-channel EEG data as reference based on the data distribution. After projecting all EEG data to the principal-component domain, high-variance components projected from the artifacts are detected and eliminated by a cutoff parameter $k$. Both ICA with ICLabel method and ASR with different cutoff parameter $k$ can generate noisy-noiseless pairs automatically. The method with best decoding performance is selected as reference method in our experiment.

\subsection{Model Training}

Assume the observed multi-channel EEG data, denoted as $X$, is composed of uncontaminated EEG signals, represented as $Y$, and signals originating from multiple sources of noise, referred to as $N$. Technically, there is currently no method that can perfectly separate pure EEG signals reflecting brain activity from the contaminated signals. In this work, we employ existing offline artifact removal methods for EEG data to produce the reference data, $\tilde{Y}_r$, which can be regarded as $Y$ based on its assumption and algorithms.

The training process is to learn the parameters of CLEEGN model, $\phi$ and we can formulate the optimization problem by following equation:

\begin{equation}
    \mathop{\arg\min}_{\phi}\mathbb{E}\{\|\tilde{Y}_r-CLEEGN_{\phi}(X)\|^{2}_{2}\}
\end{equation}

The fundamental idea of the training process is to create the optimal mapping that can transform noisy EEG data into reconstructed EEG data with minimal difference to the reference data. Usually, the EEG measurement experiments last for several minutes, it is not efficient in terms of memory usage and time consumption to use the full session EEG data directly for model training. Practically, we segment the data into multiple EEG epochs of size $(C, T)$ using the sliding windows algorithm, where $C$ is the number of electrodes of the recording device and $T$ is the number of time points in each EEG epoch. The size of the windows and the step are $T$ and $0.5T$ respectively in the sliding windows algorithm. Before the segmentation, we employ the adaptive z-normalization on each channel independently where the mean and standard deviation calculated by the noiseless EEG segment.

Considering the context of plug-and-play EEG reconstruction, we perform subject-independent training scheme where EEG data into $k$ disjoint sets by subjects. During the training process, one of the sets was left out for testing. Subjects' EEG data in the left-alone (testing) set were not involved in both training and validation. A complete experiment on a single dataset would result in $k$ different models. The artifact removal performance of a model was evaluated by using the left-alone set. The number of subjects in one set and the EEG duration available for each subject depend on the experiment setting and the dataset used.

\subsection{Datasets and Pre-processing}

In this work, both synthetic data and real EEG data are employed. In the scenario of synthetic data, we mix several independent sources to construct the multivariate time series as a simulation of the observed signals. During the experiment, a subset of the independent sources are treated as noise components. The sources in this subset are excluded while constructing the reference (noiseless) multivariate time series. Rather than use the real EEG data in early development, we can evaluate the performance and retrieve the preliminary understanding of the model behavior without the consideration of the complex spatiotemporal relationship in the EEG data. As for the experiment on real EEG dataset, to generate large-scale reference EEG data, we adopted automatic denoising methods, ICLabel and ASR, to remove artifacts and reconstruct clean waveform offline. The use of real EEG data ensures presence of artifact/noise in a natural way and thus provides a realistic evaluation for our model. The two EEG dataset are BCI-Challenge dataset (\url{https://www.kaggle.com/c/inria-BCI-Challenge}) and the MAMEM-SSVEP-II dataset (\url{https://www.mamem.eu/results/datasets/}) respectively.

\begin{figure}[ht]
    \centering
    \includegraphics[width=1.0\linewidth]{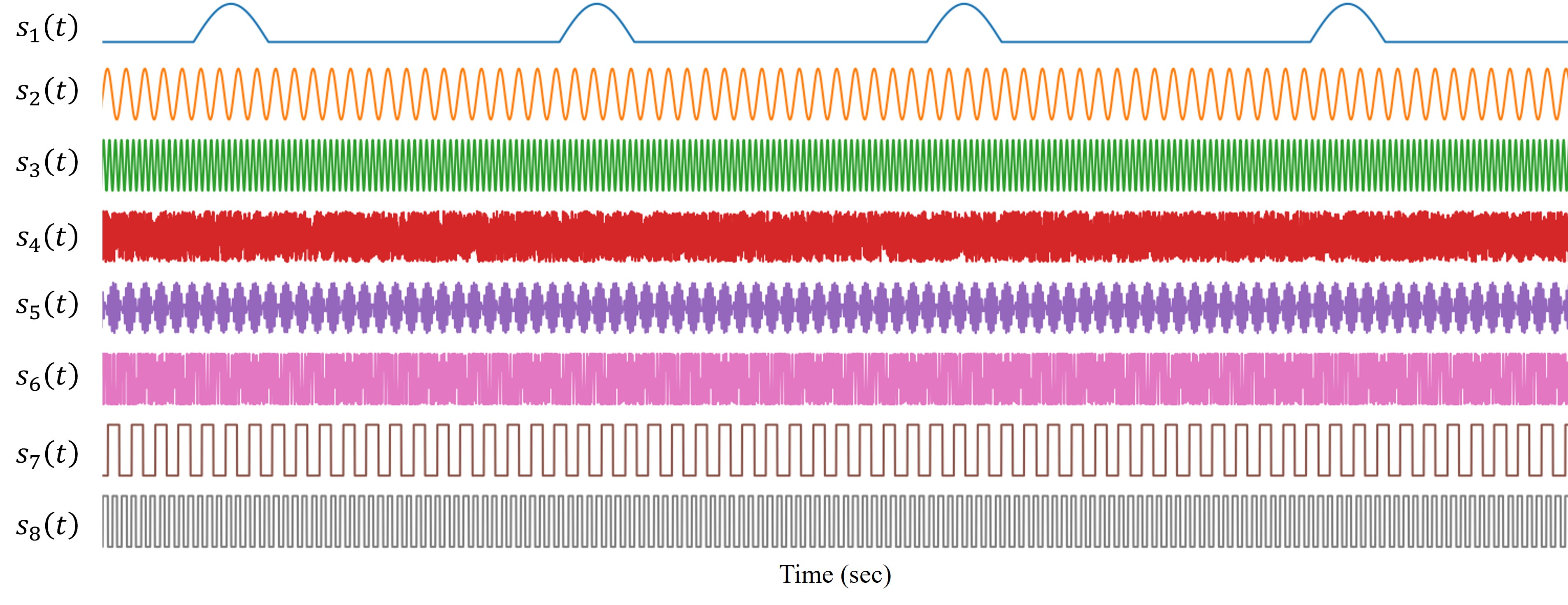}
    \caption{The waveform of the simulated sources}
    \label{fig:eight_syn_sause}
\end{figure}

\subsubsection{Synthetic Multivariate Time Series}

In the synthetic data experiment, we simulate the observed signals based on the process that used in the previous study \cite{syngenref}. Eight temporal signal sources, $s_1(t)$ to $s_8(t)$, are generated, each with a duration of 100 seconds and a sample frequency of 1000 Hz. The waveform of eight simulated sources are presented in Figure \ref{fig:eight_syn_sause}. $s_1(t)$ simulates the irregular large amplitude noise, several peak signals are positioned randomly in a zero-magnitude signal. $s_2(t)$ and $s_3(t)$ are typical periodic sources, specifically 8 Hz and 25 Hz sinusoidal waves. $s_4(t)$ is a random series generated based on uniform distribution. $s_5(t)$ and $s_6(t)$ are composite periodic function with equations $\mathop{\sin}(400t)\mathop{\cos}(30t)$ and $\mathop{\cos}(400t + 10\mathop{\sin}(90t))$. The remaining two source simulate two square wave with different densities.

\begin{figure}[ht]
    \centering
    \includegraphics[width=1.0\linewidth]{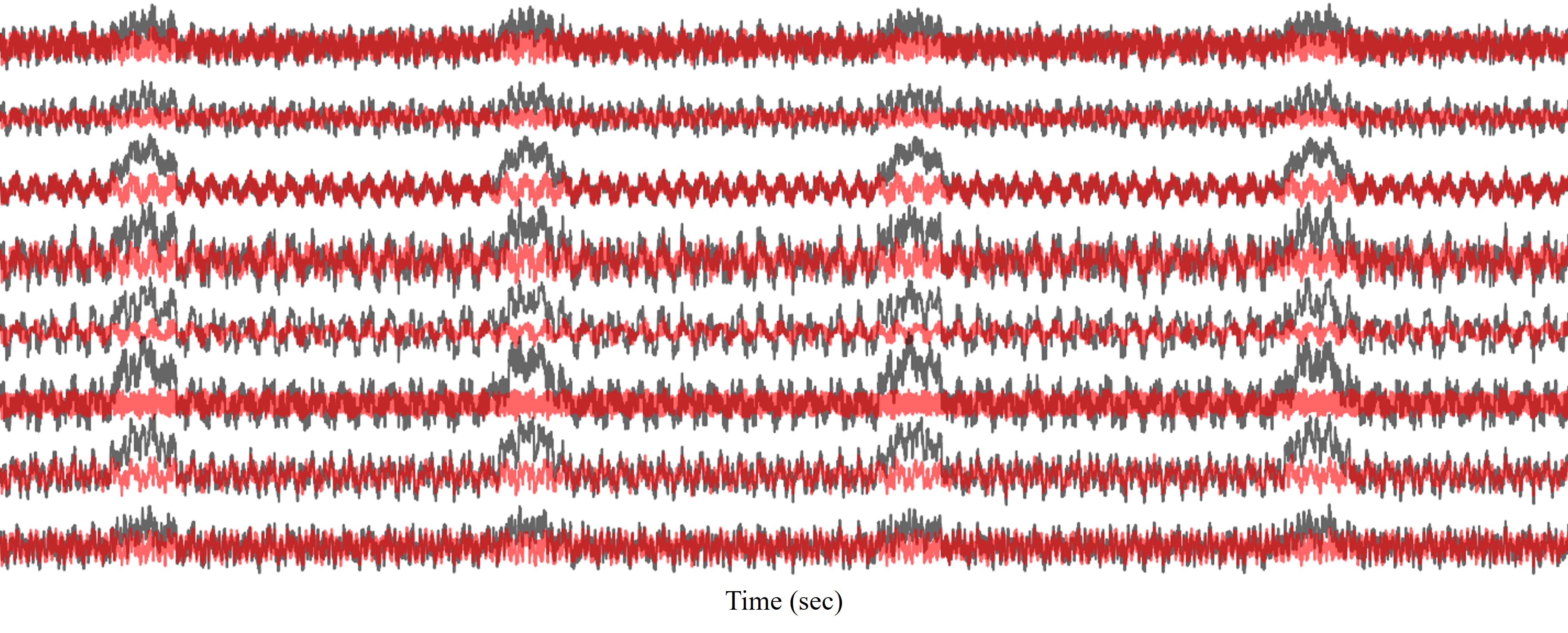}
    \caption{An example of the synthesized signal mixture (black) and the corresponding reference signals (red)}
    \label{fig:eight_syn_mix057}
\end{figure}

In this experiment, we consider three scenarios with different distributions of desirable sources, $\phi_d$, and undesirable sources, $\phi_u$: scenario-1) $\phi_d=\{s_2(t),s_3(t),$$s_4(t),s_5(t),s_6(t),s_7(t),s_8(t)\}$, $\phi_u=\{s_1(t)\}$, scenario-2) $\phi_d=\{s_2(t),s_3(t),s_4(t),s_5(t),s_6(t)\}$, $\phi_u=\{s_1(t),$ $s_7(t),s_8(t)\}$, scenario-3) $\phi_d=\{s_2(t),s_3(t)\}$, $\phi_u=\{s_1(t),$ $s_4(t),s_5(t),s_6(t),s_7(t),s_8(t)\}$.

To synthesize a signal mixture, the eight source signals are formulated into a source matrix $S=[s_1(t),s_2(t),\cdots,s_8(t)]^T$. Next, a mixing matrix $A\in\mathbb{R}^{8\times8}$ is generated randomly by uniform distribution, $A\sim U(0,1)$. The observed signal mixture $X$ is defined as
\begin{equation}
    X=AS+\epsilon.
\end{equation}
The $\epsilon=[n_1(t),n_2(t),\cdots,n_8(t)]^T$ is used to simulate the subtle perturbation present in the signals, where $n_i(t)\sim\mathcal{N}(0,\,0.01),\,i=1,2,\cdots,8$. 

For a signal mixture, we generate the corresponding reference signals by excluding the undesired sources and projecting only the desirable sources with the same mixing matrix. Take the scenario-2 as an example, the noiseless source matrix can be formulated as $S^*=[z(t),s_2(t),\cdots,s_6(t),z(t),z(t)]^T$, where $z(t)$ represents a zero-magnitude signal and the mixture reference signal matrix becomes
\begin{equation}
    Y=AS^*+\epsilon
\end{equation}
An example of the observed signal mixture and the corresponding reference signals in the scenario-2 is shown in Figure \ref{fig:eight_syn_mix057}.

\subsubsection{BCI-Challenge}

Error-related negativity (ERN) \cite{ERNdesc} can be categorized as a type of event-related potential (ERP) \cite{ERPdesc}, which emerges after an erroneous or abnormal event perceived by the subject. One characteristic of the feedback ERN is a relatively large negative amplitude approximately 350 ms and a positive amplitude approximately 500 ms after visual feedback triggered by the error event. In this work, we mainly use a well-studied EEG dataset from the BCI-Challenge competition hosted by Kaggle to evaluate the artifact removal effectiveness. This dataset includes EEG recordings of 26 subjects (16 subjects labeled and 10 subjects unlabeled) that participated in a P300 speller task \cite{krusienski2008toward}.

P300 speller is a well-known BCI system that develops a typing application through P300 response evoked potential. The ERN experiment was conducted under the assumption that the ERN occurred if the subject received incorrect prediction (feedback) from the P300 speller. The objective of the competition was to improve the P300 speller performance by implementing error correction through ERN potentials. We used the 16 subjects with labeled data of which the sampling rate is 200 Hz initially with 56 passive Ag/AgCl EEG sensors.

In the interest of increasing the usability of EEG data, we applied some pre-processing procedures to each EEG recording. The EEG data were down-sampled to 128 Hz and re-referenced by the common average reference (CAR) method to eliminate common-mode noise and to zero-center the data. Each recording was band-pass filtered to 1-40 Hz through the FIR filter implemented by EEGLAB \cite{delorme2004eeglab} to remove DC drifting. During the EEG decoding evaluation, we epoch the EEG signal in [0, 1.25] second interval to obtain correct and erroneous feedback.

\subsubsection{MAMEM-SSVEP-II - Steady state visually evoked potential (SSVEP)}

Steady state visually evoked potential (SSVEP) \cite{iscan2018ssvepbci} can also be categorized as a kind of ERP. The potentials are characterized as periodic potential induced by rapidly repetitive visual stimulation. The SSVEP is composed of several discrete frequency components, which consist of the fundamental frequency of the visual stimulus as well as its harmonics. To investigate the generalization ability of the model, we use "EEG SSVEP Dataset II" from Multimedia Authoring \& Management using your Eyes \& Mind (MAMEM). The dataset includes EEG data from 11 subjects and consists of five different frequencies (6.66, 7.50, 8.57, 10.00, and 12.00 Hz). Each subject was recorded in five sessions and each session included 25 trials (5 trials for each class). The data used a 256-channel HydroCel Geodesic Sensor Net (HCGSN) and captured the signals with a sampling rate of 250 Hz.

We selected a common montage of 20 channels across subjects to train the reconstruction model. Every recording was down-sampled to 125 Hz, re-referenced by the common average reference (CAR) method, and band-pass filtered to 1-40 Hz. We epoch EEG signals in [1, 5] second interval for each event recorded timestamp. The first second was discarded under the consideration of a reaction delay of the subject.

\subsection{Performance Evaluation and Metrics}

We design a simulated online reconstruction process that performs artifact removal on a multi-channel EEG data from a new-coming subject based on a trained EEG reconstruction model without any requirement of training/calibration. 
A simulated online stream of raw EEG data accessed from above-mention datasets were epoched into 4-s segments with a 1-s stride. 
Each segment of the raw signal is processed sequentially, and we discard the edges since the employed zero-padding in the model causes attenuation in amplitude on the sides of the outcome. All of the generated results were further concatenated to form a full session reconstructed EEG data.

In this work, we compare the performance of CLEEGN against four neural-network based methods, 1D-ResCNN \cite{SUN2020108}, IC-U-Net \cite{CHUANG2022119586} and the simple CNN (SCNN) structure and RNN structure proposed in EEGdenoiseNet \cite{EEGdenoiseNet}. Both the subjective evaluation on the visualized results and several objective assessment metrics are used to assess the performance of each method. The objective assessment metrics of the synthetic data are listed in Table \ref{tb:metric_all}.

\begin{table}[h]
\caption{Performance assessment metrics for synthetic EEG data reconstruction.}
\label{tb:metric_all}
\centering
\resizebox{\linewidth}{!}{
\begin{tabular}{m{1.0\linewidth}}
    \hline
    Metric \\
    \hline
    \\
    $MSE=\mathbb{E}\{\|EEG_{ref}-EEG_{est}\|^{2}_{2}\}$ \\
    \\
    Mean square error estimates the fitness in the time domain. Better fitness of the model generally corresponds to lower MSE value. The $EEG_{ref}$ denotes the reference EEG data and $EEG_{est}$ is the estimated EEG data. \\
    \hline
    \\
    $MPCC=\sqrt[C]{\prod_{k=1}^{C}corr(EEG^{k}_{ref},EEG^{k}_{est})}$ \\
    \\
    Mean Pearson correlation coefficient estimates the linear relationship between the reference and the estimated EEG data in the time domain. The $corr(\cdot)$ in the formula is the Pearson correlation coefficient. The $EEG^{k}_{ref}$, $EEG^{k}_{est}$ denote the $k$-th channel from the $C$-channel reference data and reconstructed outcome respectively. \\
    \hline
    \\
    $MAE_{PSD}=\frac{1}{C}\sum_{k=1}^{C}\frac{\|PSD^{k}_{ref}-PSD^{k}_{est}\|_{1}}{L}$ \\
    \\
    Mean absolute error (MAE) of the power spectral density (PSD) is utilized to analyze the spectrum distortion. The $PSD^{k}_{ref}$, $PSD^{k}_{est}$ represent the power spectral density of the $k$-th channel of the $C$-channel reference signals and reconstructed signals and $L$ denotes the number of frequency bins. The power spectral density is estimated by discrete Fourier transformation. \\
    \hline
    \\
    $MPCC_{IC}=\sqrt[M]{\prod_{i=1}^{M}\mathop{\max_{j=1,\cdots,M}}corr(S^{i}_{ref},S^{j}_{est})}$ \\
    \\
    Mean Pearson correlation coefficient of the decomposition estimates the information gain after the reconstruction. The fastICA algorithm \cite{fastica} is employed to decomposed the reconstructed EEG data. The $S^{i}_{ref}$, $S^{i}_{est}$ denote the $i$-th component of the $M$-dimensional decomposition result of the reference data and reconstructed outcome respectively. \\
    \hline
\end{tabular}
}
\end{table}

To objectively assess the data quality of reconstructed real EEG, we adopt the decoding performance to measure the presence of informative EEG activity using the two labeled real EEG dataset. We herein use EEGNet \cite{lawhern2018eegnet}, a compact CNN for end-to-end EEG decoding as the classifier to decode the labeled EEG data in our study. The classification performance for MAMEM-SSVEP-II dataset is estimated by the Top-1 accuracy. Considering that BCI-Challenge is an unbalanced binary dataset, we estimate the performance using the score of Area Under the Curve (AUC). 
The EEG epochs are divided into a training, a validation, and a test set with consistent ratio between classes for each subject.
We perform 20 repeated runs with shuffled data and acquire the average performance of each decoding performance evaluation.

\section{Results and Discussion}

We evaluate CLEEGN's performance against leading baseline methods across three scenarios: synthetic EEG data and two real EEG datasets. In addition, we conduct experiments for assessing the effect of reference data selection, session length, and training data size on the performance. Furthermore, we present the visualization of latent features to elucidate the behavior of the model.

\subsection{Validation on the Synthetic Data}

\begin{table*}[ht]
\caption{
Performance evaluation and comparison on the synthetic EEG data.
}
\label{tb:syn_evaluation}
\centering
\resizebox{\linewidth}{!}{\begin{tabular}{c|cccc|cccc|cccc}
    \hline
    \multirow{2}{*}{Method} & \multicolumn{4}{c|}{scenario-1} & \multicolumn{4}{c|}{scenario-2} & \multicolumn{4}{c}{scenario-3} \\
    \cline{2-13}
    & $MSE\,\downarrow$ & $MPCC\,\uparrow$ & $MAE_{PSD} (10^{-5})\,\downarrow$ & $MPCC_{IC}\,\uparrow$ &
    $MSE\,\downarrow$ & $MPCC\,\uparrow$ & $MAE_{PSD} (10^{-5})\,\downarrow$ & $MPCC_{IC}\,\uparrow$ &
    $MSE\,\downarrow$ & $MPCC\,\uparrow$ & $MAE_{PSD} (10^{-5})\,\downarrow$ & $MPCC_{IC}\,\uparrow$ \\
    \hline
    Raw data (noisy) & 1.0289 & 0.7502 & 1.7407 & 0.8382 &
        1.6738 & 0.4802 & 3.5312 & 0.4763 &
        1.9376 & 0.2472 & 10.7059 & 0.2669 \\
    \hline
    1D-ResCNN & 0.4099 & 0.7811 & 8.5632 & 0.4962 &
        0.3231 & 0.5731 & 8.4124 & 0.4699 &
        0.0965 & 0.6560 & 1.5747 & 0.5218 \\
    RNN & 0.2039 & 0.8956 & 6.0912 & 0.7909 &
        0.1552 & 0.8372 & 4.6724 & 0.6553 &
        0.0659 & 0.7632 & 1.0395 & 0.8265 \\
    SCNN & 0.2592 & 0.8657 & 6.3754 & 0.7515 &
        0.1947 & 0.7819 & 5.4097 & 0.7092 &
        0.2048 & 0.4800 & 5.8007 & 0.4061 \\
    IC-U-Net & 0.1752 & 0.9369 & 3.7872 & 0.9097 &
        0.0879 & 0.9267 & 3.4064 & 0.9241 &
        0.0118 & 0.9745 & 1.4577 & 0.9855 \\
    CLEEGN & \textbf{0.0082} & \textbf{0.9956} & \textbf{1.0112} & \textbf{0.9784} &
        \textbf{0.0054} & \textbf{0.9943} & \textbf{0.9154} & \textbf{0.9840} &
        \textbf{0.0024} & \textbf{0.9852} & \textbf{0.7472} & \textbf{0.9910} \\
    \hline
\end{tabular}}
\end{table*}

\begin{figure}[ht]
    \centering
    \includegraphics[width=1.0\linewidth]{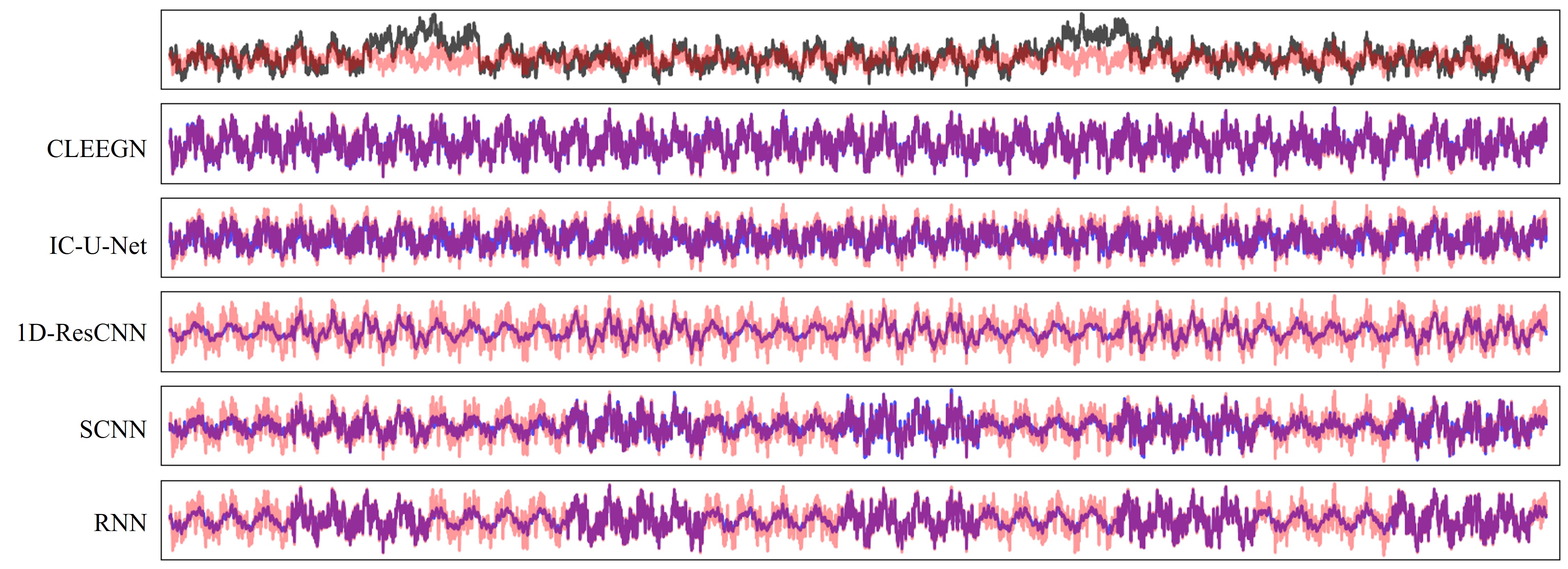}
    \caption{An experimental result waveform for eliminating noise sources from an synthetic signal under scenario-2. The first raw illustrates the raw signal (black) without suppressing noise and the ground truth signal (red). The following from top to bottom, the waveform colored in blue are denoised by our method, IC-U-Net, 1D-ResCNN, SCNN and RNN.}
    \label{fig:syn_ban057_ch3}
\end{figure}

\begin{figure}[ht]
    \centering
    \includegraphics[width=1.0\linewidth]{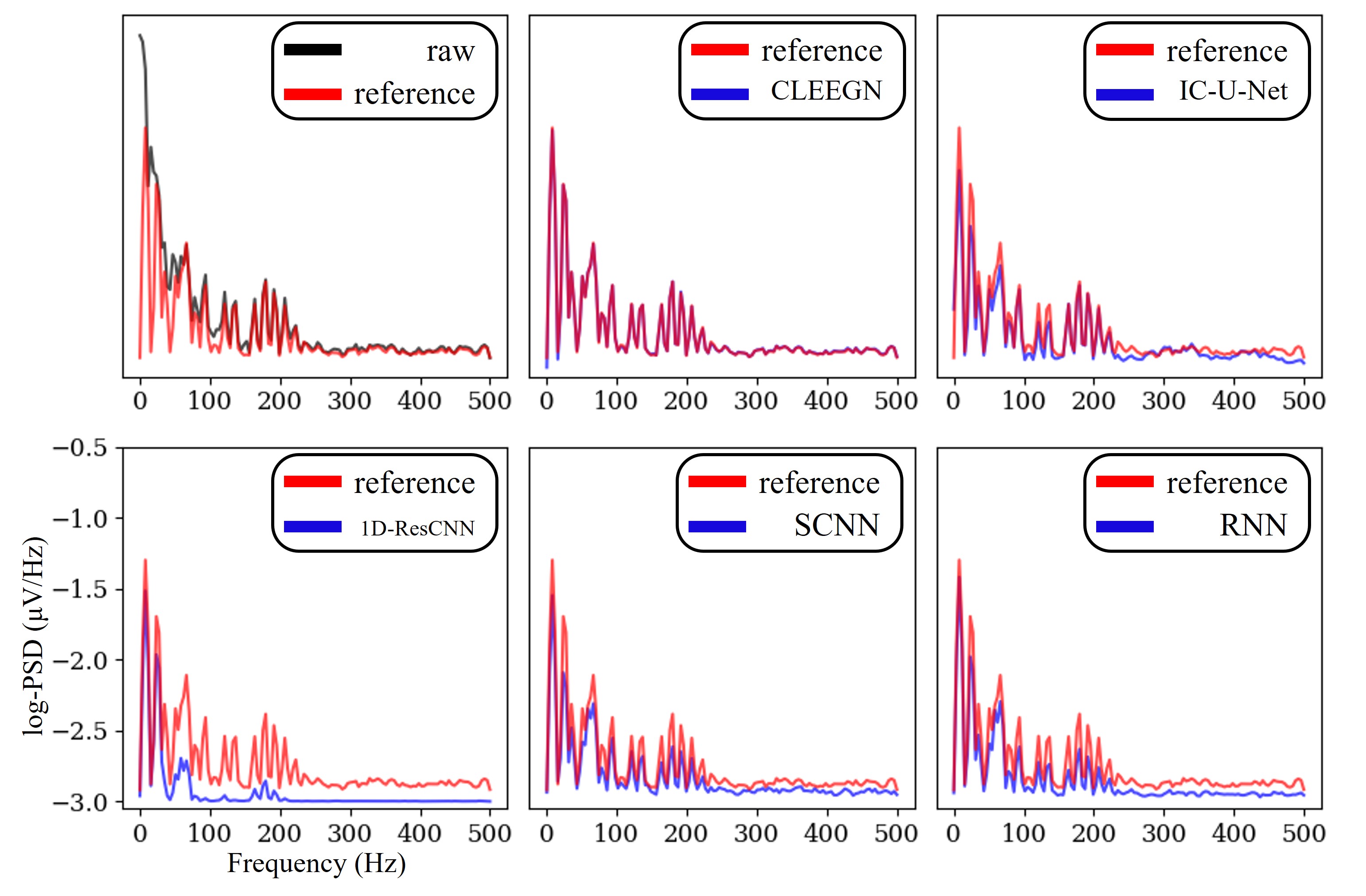}
    \caption{
    Reconstruction of synthetic data in power spectral density (PSD) using various methods under scenario-2. The top-left subfigure compares the PSD of the raw signal mixture (gray) with the reference signal (red). The remaining subfigures compare the reconstruction signals (blue) from different methods to the reference signals (red).
    }
    \label{fig:syn_ban057_ch2_PSD}
\end{figure}

\begin{figure}[ht]
    \centering
    \includegraphics[width=1.0\linewidth]{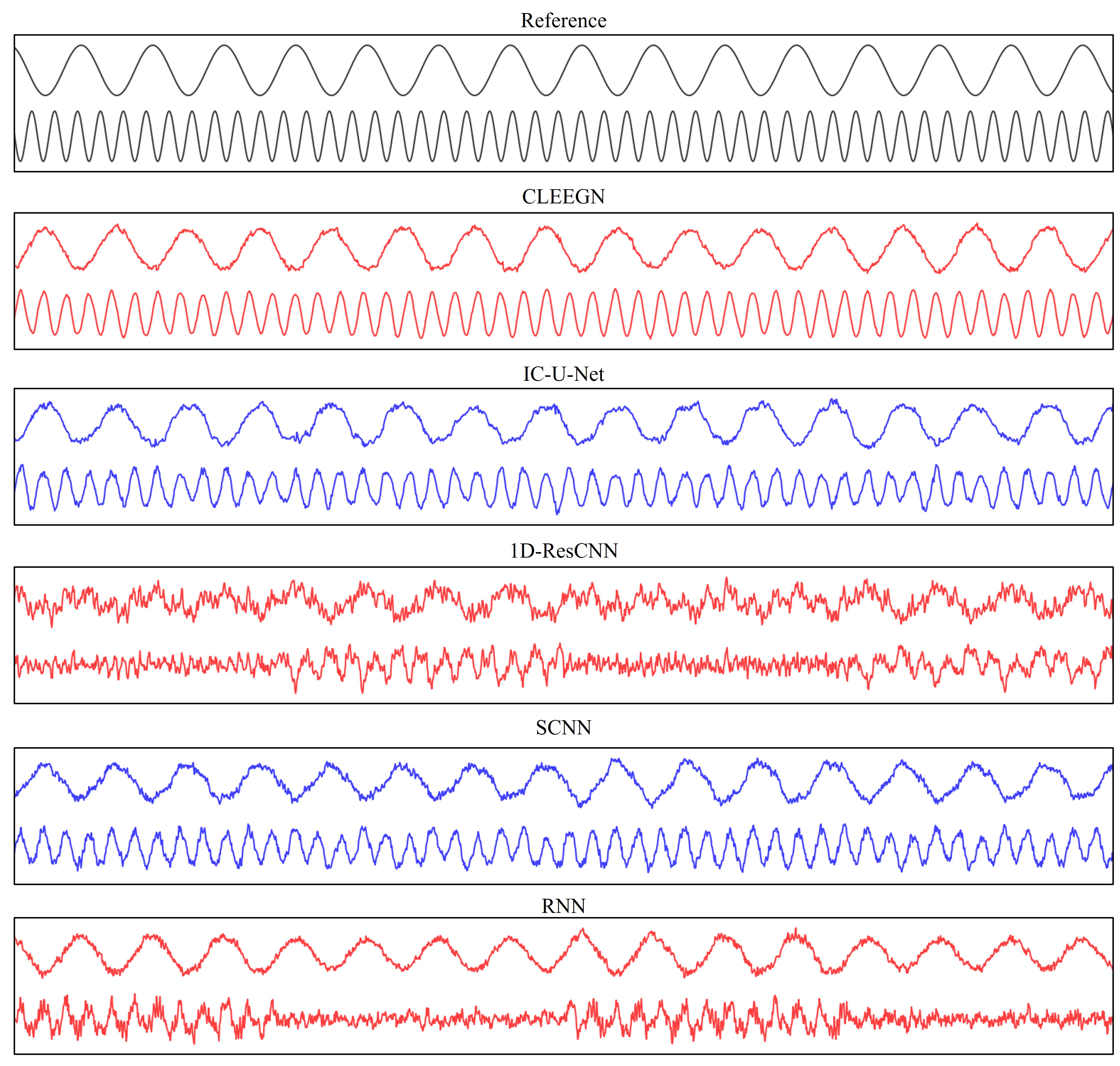}
    \caption{Two most similar component to 8Hz and 25Hz decomposed by fastICA of the reference and the reconstructed result from each method under the scenario-2.}
    \label{fig:syn_ban057_fastICA}
\end{figure}

In Figure \ref{fig:syn_ban057_ch3},\ref{fig:syn_ban057_ch2_PSD},\ref{fig:syn_ban057_fastICA}, we compare the outputs of each denoising network method with the reference method on the synthetic scenario-2 in terms of time and frequency domains as well as the feature representation.

The Figure \ref{fig:syn_ban057_ch3} demonstrates a 5 seconds segment of the reconstructed results from each method. The signal colored in black is the raw data without adopting any noise cancellation method. The visualized result shows that all neural network-based methods, including the proposed method, have the ability to alleviate the influence of high amplitude noise. Compared to 1D-ResCNN, SCNN and RNN methods, the characteristic of the reconstruction from CLEEGN and IC-U-Net are similar to the reference method. We can also observe that the reconstructed result generated by the CLEEGN fit the reference signal more closely in comparison with IC-U-Net.

Figure \ref{fig:syn_ban057_ch2_PSD} shows the difference of the reconstructed results from various methods under the perspective of frequency domain. According to the visualization, the signal reconstructed by 1D-ResCNN has the largest distortion in the power spectrum. Though the SCNN and RNN perform better than 1D-ResCNN, this method still falls slights short in reconstructing each spectral peaks and valleys. In comparison, IC-U-Net and our proposed method exhibit a better fitness to the power spectrum of the reference signal. Especially for CLEEGN, we can observe that our approach results in less deformation than other methods in the reconstruction of spectral power.

\begin{figure*}[ht]
    \centering
    \includegraphics[width=.9\linewidth]{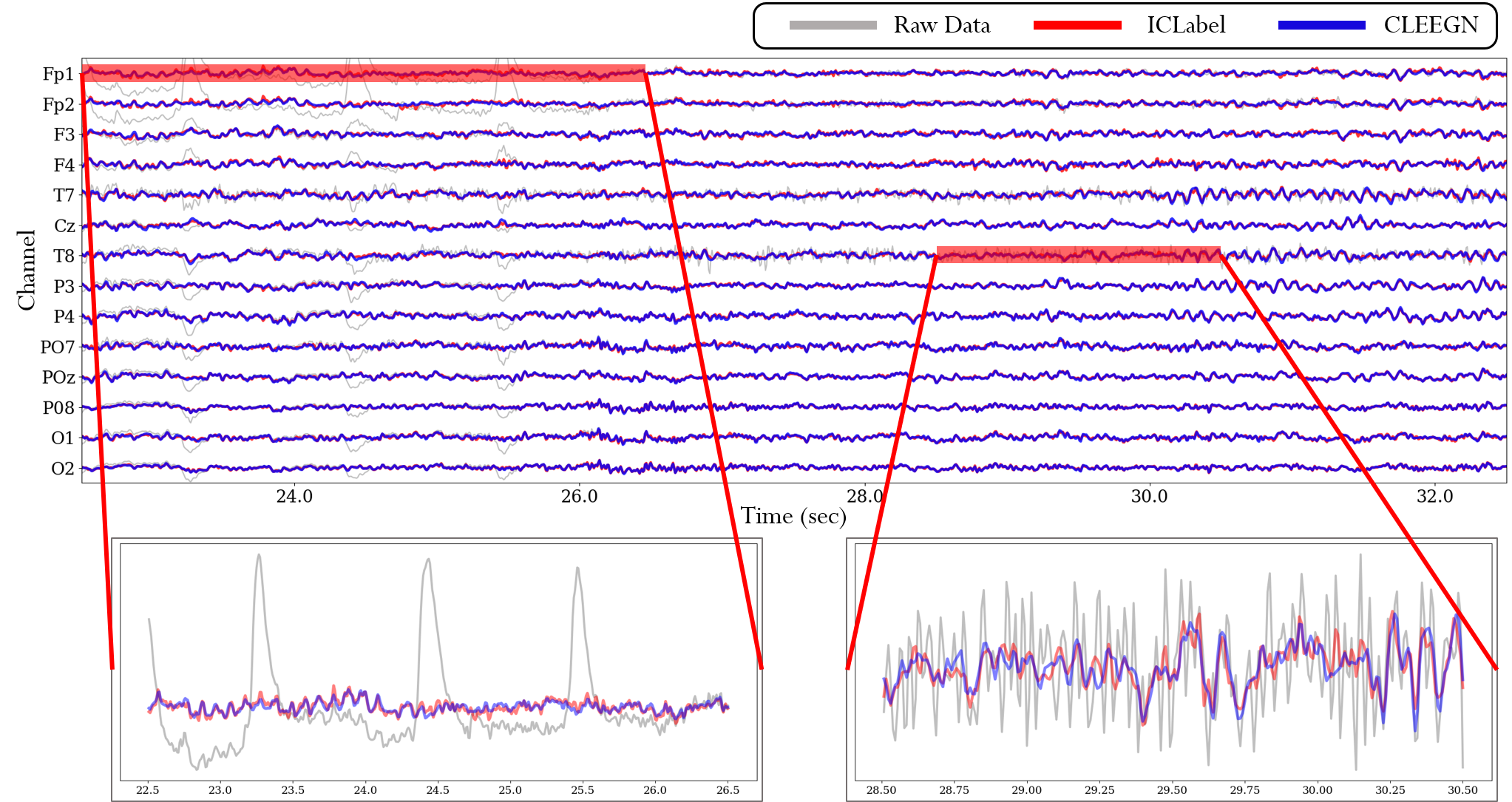}
    \caption{A 10 seconds waveform demonstration of raw EEG (gray), reference EEG (red) through ICA+ICLabel method and reconstructed results (blue) by CLEEGN.}
    \label{fig:bigwave}
\end{figure*}

Next, we further examine the reconstruction results through component analysis.
In Figure \ref{fig:syn_ban057_fastICA}, we decompose the reconstruction from each method through the fastICA algorithm \cite{fastica} and illustrate two independent components (ICs) that are most similar to the 8Hz and 25Hz sinusoidal waves. The similarity is estimated by the absolute Pearson correlation coefficient. From top to bottom, the results are reference synthetic data, CLEEGN, IC-U-Net, 1D-ResCNN, SCNN and RNN, respectively. Generally, the outcome presented by RNN is better than the one from the 1D-ResCNN. From the result of RNN, we can see more obvious 8Hz oscillations from the first decomposed component. Nevertheless, its second decomposition result presents a combined pattern of multiple sinusoidal waves instead of a distinguishable 25 Hz waveform. From the visualized result, the difference is less obviously between the decomposition between our proposed method, IC-U-Net and SCNN and the reference. In terms of the visualization perspective, CLEEGN demonstrate the least distortion on the decomposed result. It implies that CLEEGN have the superior ability of preserving desired information while removing the noise from the contaminated data.

Table \ref{tb:syn_evaluation} summarizes the objective estimation on the synthetic dataset under three different scenarios. In each scenario, the reconstructed signals from our method exhibit minimal Mean Squared Error (MSE) with the reference compared to other methods. Furthermore, the outcomes of CLEEGN demonstrate the best alignment with the reference signals in the time domain, resulting in the highest Pearson Correlation Coefficient (MPCC) value. In spectral evaluation, the Mean Absolute Error (MAE) estimation on the power spectral density shows that CLEEGN introduces less distortion than other reconstruction methods in the frequency domain.
On the other hand, CLEEGN provides the highest statistical correlation with the reference after decomposing the signals using the fastICA algorithm, indicating its high information-preserving capability during the reconstruction.

\subsection{Visual Inspection of EEG Reconstruction Using CLEEGN}

In the experiments of real EEG data reconstruction, we first inspect the reconstruction waveforms on visually-discernible artifacts such as eye blinks and muscular artifacts \cite{Sazgar2019}. Figure \ref{fig:bigwave} is a 14 selected channels, 10 seconds EEG recording from the BCI-Challenge dataset. Comparing the reference signals generated by the ICLabel method (red) and reconstructed results (blue), the waveform reconstructed by CLEEGN highly overlaps to the reference, which imply the great fitting capability of the CLEEGN structure. Inspecting the raw data plotted in light gray, we can find two types of conspicuous artifacts in this time session. First, there are few high-amplitude ocular artifacts that are phenomenal on Fp1, Fp2, etc around 0 to 4 seconds. The second one is the muscular artifacts featuring high-frequency ($\sim$24-30 Hz in the $\beta$ band) observed on the T8 channel near the right ear in the 6-8 seconds of this time session. We can see that both the ICLabel method and CLEEGN can improve the signal quality by eliminating these kinds of artifacts.

\subsection{Performance Comparison on Real EEG Data}

\begin{figure}[t]
    \centering
    \includegraphics[width=.9\linewidth]{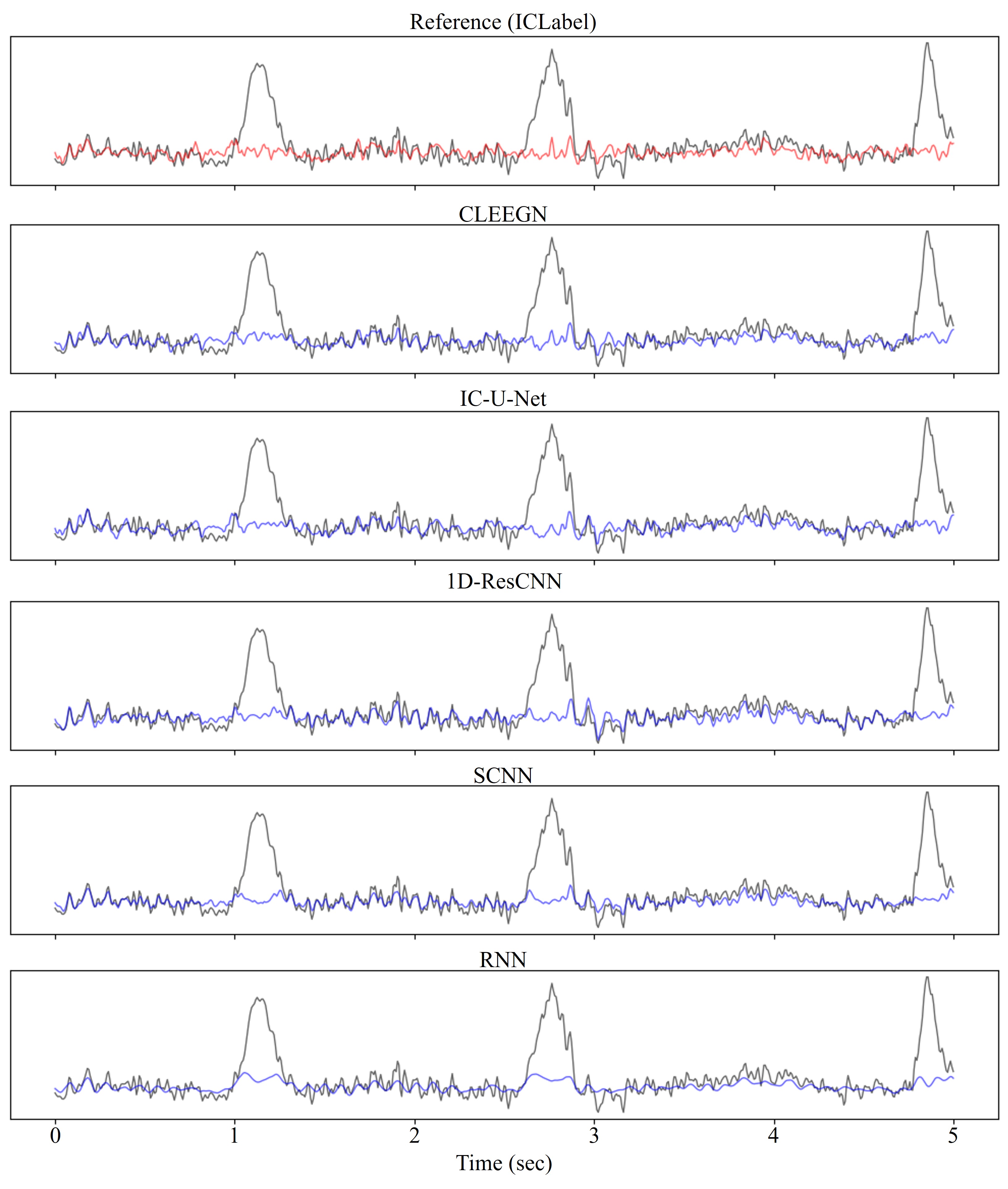}
    \caption{The comparison of each artifact removal method on 5 seconds waveform from Fp1 channel with significant EOG artifacts in BCI-Challenge dataset.}
    \label{fig:vis_Fp1_s2_EOG}
\end{figure}

Figure \ref{fig:vis_Fp1_s2_EOG} illustrates the outcome of eliminating eye-related artifacts from the EEG signal in the BCI-Challenge dataset. In each graph, the gray line represents a 5-second raw EEG signal from the frontal channel (Fp1), displaying three noticeable EOG artifacts. From top to bottom, the artifacts are suppressed by ICA with ICLabel which is treated as reference signal, CLEEGN, IC-U-Net, 1D-ResCNN, SCNN and RNN. According to the waveform visualization, the reconstructed signal in the noiseless period within each method is quite similar, except for the distortion of high-frequency patterns when using RNN model. As for the reconstruction of eye-related artifact contamination, CLEEGN, IC-U-Net and 1D-ResCNN presents comparable performance in fitting the reference signals in this visual inspection.

\begin{table}[h]
\caption{Overall performance over all subjects in the BCI-Challenge dataset.}
\label{tb:arnn_ern}
\centering
\resizebox{\linewidth}{!}{\begin{tabular}{llll}
    \hline
    Method & MSE $\downarrow$ & \#params $\downarrow$ & AUC $\uparrow$ \\
    \hline
    ASR-4 & -- & -- & 0.6300$\pm$0.0249 \\
    ASR-8 & -- & -- & 0.6741$\pm$0.0277 \\
    ASR-16 & -- & -- & 0.7098$\pm$0.0273 \\
    ASR-32 & -- & -- & 0.7263$\pm$0.0281 \\
    ICA + ICLabel (reference) & -- & -- & 0.7398$\pm$0.0273 \\
    Raw data (contaminated) & -- & -- & 0.5578$\pm$0.0100 \\
    \hline
    1D-ResCNN & 6.7147$\pm$0.5025 & 325891 & 0.6840$\pm$0.0306 \\
    RNN       & 10.6686$\pm$0.9257 & 787984 & 0.6872$\pm$0.0292 \\
    SCNN      & 7.7823$\pm$0.6403 & 16815552 & 0.6872$\pm$0.0259 \\
    IC-U-Net  & 5.2086$\pm$0.4204 & 2683192 & 0.7251$\pm$0.0259 \\
    CLEEGN    & 3.5984$\pm$0.2538 & \textbf{220755} & \textbf{0.7494$\pm$0.0264} \\
    \hline
\end{tabular}}
\end{table}

\begin{figure}[ht]
    \centering
    \includegraphics[width=.9\linewidth]{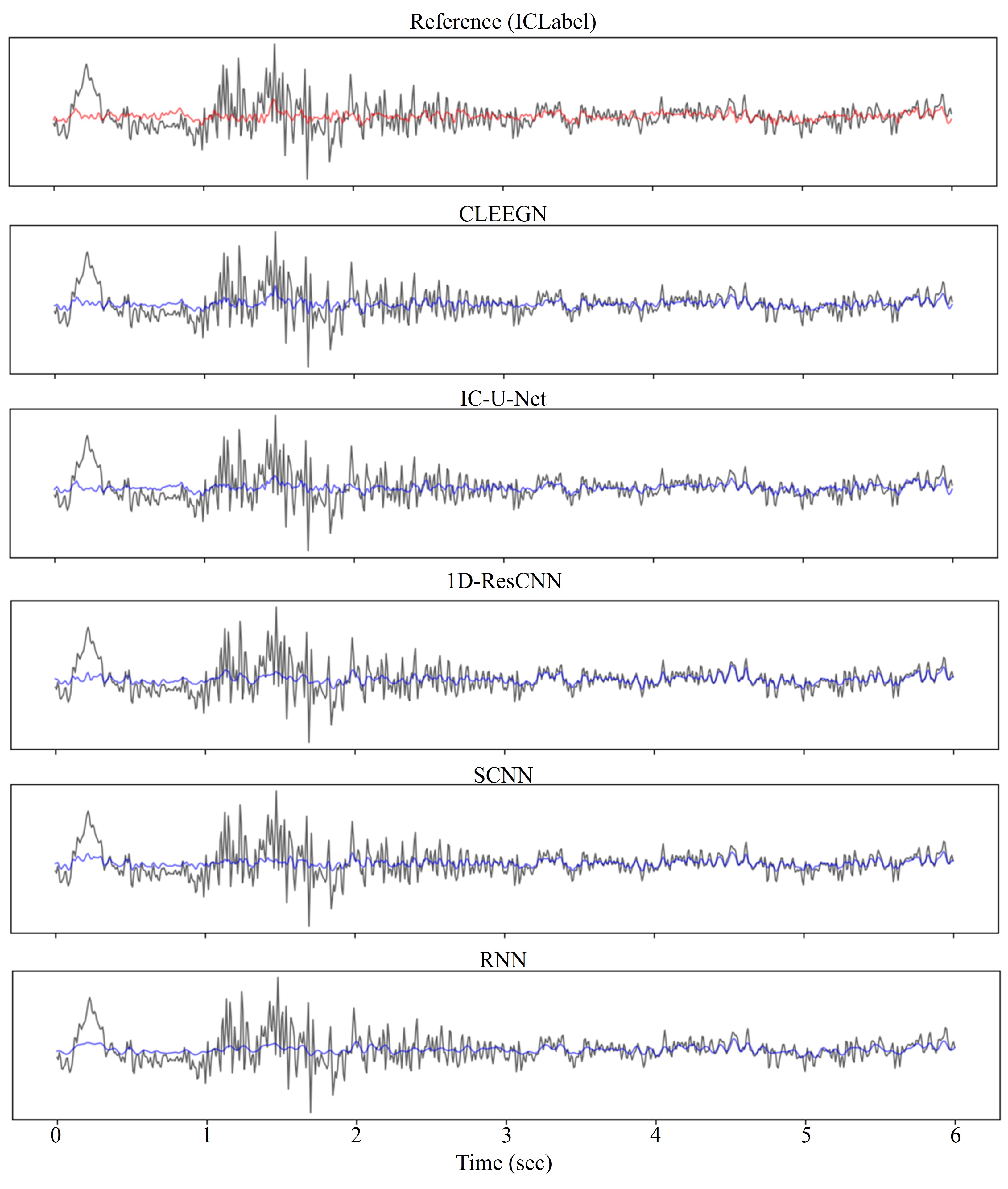}
    \caption{The comparison of each artifact removal method on 6 seconds waveform from F3 channel with significant EMG artifacts in BCI-Challenge dataset.}
    \label{fig:vis_F3_s2_EMG}
\end{figure}

Figure \ref{fig:vis_F3_s2_EMG} presents the result of removing muscular artifact from the EEG signal in BCI-Challenge dataset. In the selected period, the raw EEG signal (gray) from F3 channel with an eye blink followed by muscular artifact from 1-3 s. The signal reconstructed by CLEEGN and IC-U-Net are both highly similar to the reference than other method, particularly on the positive potential around 1.5 s. From the visual perspective, SCNN is considered to have better reconstruction ability than 1D-ResCNN in the EMG contaminated period since the result of 1D-ResCNN is overly smooth. The visualization result shows that RNN short on reserving high frequency oscillation.

In addition the analyze from the visualization, Table \ref{tb:arnn_ern} shows the objective evaluation of each neural network-based artifact removal method in the same dataset. The first and the second row are the statistic of the ICLabel and the noisy raw data, which represents the baseline and the state before processing. The reconstructed signals from our proposed method has the best fitness to the reference signals in the time domain, which provides the minimum MSE value. Additionally, CLEEGN provides the highest AUC score in the EEG classification. It implies that the reconstructed result still preserves critical features related to the brain activity. We can also observe that our model has the least parameters in the structure.

\begin{figure}[ht]
    \centering
    \includegraphics[width=1.0\linewidth]{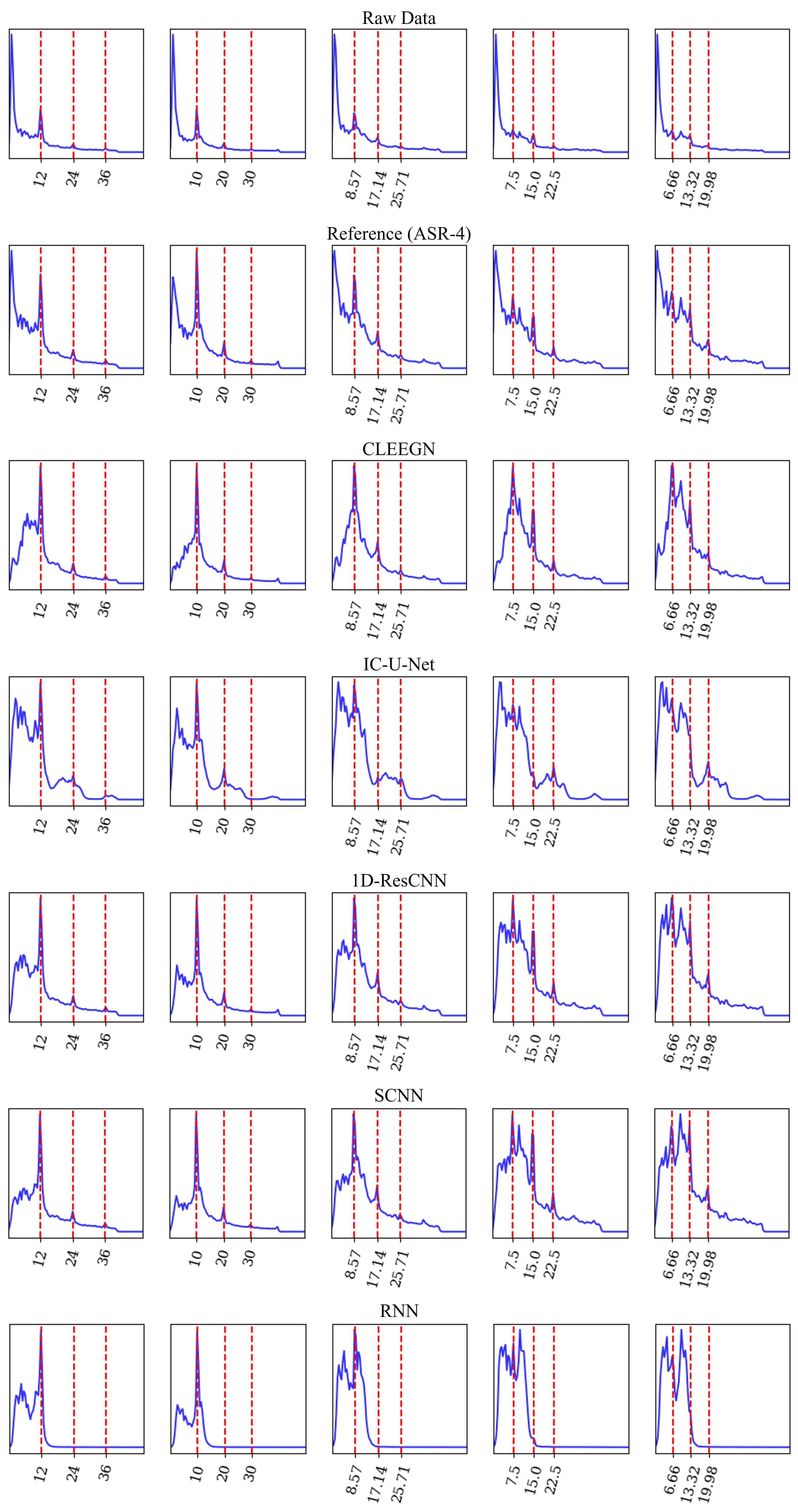}
    \caption{Accumulating power spectrum density of channel Oz with 12 Hz, 10 Hz, 8.57 Hz, 7.5 Hz and 6.66 Hz stimulus of each artifact removal method.}
    \label{fig:ssvep_class_psd}
\end{figure}

\begin{figure*}[ht]
    \centering
    \includegraphics[width=1.0\linewidth]{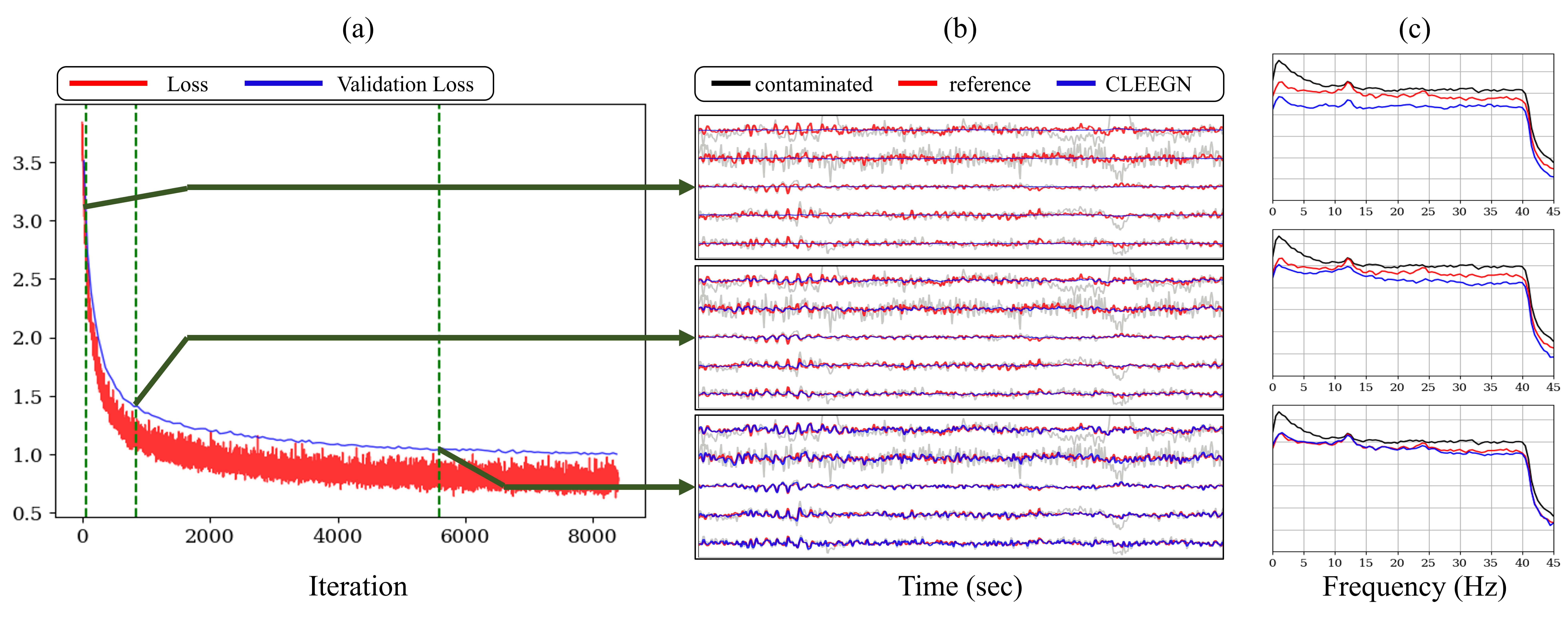}
    \caption{Reconstructed EEG signals by CLEEGN across training steps.}
    \label{fig:inter_vis_Adv}
\end{figure*}

In Figure \ref{fig:ssvep_class_psd}, from top to bottom, present the spectral analyze of raw data, reference (ASR-4), CLEEGN, IC-U-Net, 1D-ResCNN, SCNN and RNN on the MAMEM-SSVEP-II dataset. We plot the accumulating power spectrum of the trials with the same frequency stimulus on the Oz channel over each subject and session in this dataset. The red vertical dashed lines in each diagram represent the fundamental frequency stimulus and its harmonics which correspond to the values on the x-coordinate. In general, the components with these specific frequencies are induced by the visual stimulation in the experiment, and we except to observe a higher energy response at these frequencies in the spectrum. Without any noise suppression algorithm, the peaks at these certain frequencies are not noticeable since the low frequency artifacts overwhelm the presentation of the brain signals. In the experiment on MAMEM-SSVEP-II dataset, since the duration of event-marked epochs constitute a significant portion of each EEG session, the induced potentials are highly classified as artifact by the ICLabel. We choose ASR with cutoff parameter $k$ is 4 as the reference methods instead of the ICA with ICLabel method. The accumulating spectrum of the reference signals show that the effect of low frequency components are suppressed and we can observe the higher frequency responses at the frequency stimulus and its harmonics. Compares with the reference, the accumulating spectrum by neural network-based methods show that the response in the low frequency bands decrease significantly. From the perspective on the spectral analyze, the response in certain frequency bands attenuate in the outcomes of IC-U-Net, and the these spectra clearly show that RNN short on reconstructing high frequency components. The spectra from the 1D-ResCNN, SCNN and CLEEGN have high spectral resolution as well as the strong interoperability. We can see the distinguishable peaks at the stimulus and the harmonious frequencies from the accumulating spectrum by our proposed method, which indicates that our network structure is capable of preserving the spectral information in the EEG data while reconstructing the noiseless signals.

\begin{table}[h]
\caption{Overall performance over all subjects in the MAMEM-SSVEP-II dataset.}
\label{tb:arnn_ssvep}
\centering
\resizebox{\linewidth}{!}{\begin{tabular}{llll}
    \hline
    Method & MSE & \#params $\downarrow$ & Accuracy $\uparrow$ \\
    \hline
    ASR-4 (reference) & -- & -- & 0.5823$\pm$0.0194 \\
    ASR-8 & -- & -- & 0.5815$\pm$0.0228 \\
    ASR-16 & -- & -- & 0.5775$\pm$0.0229 \\
    ASR-32 & -- & -- & 0.5729$\pm$0.0217 \\
    ICA+ICLabel & -- & -- & 0.4210$\pm$0.0232 \\
    Raw data (contaminated) & -- & -- & 0.4893$\pm$0.0304 \\
    \hline
    IC-U-Net  & 1.9188$\pm$0.4048 & 2664724 & 0.3289$\pm$0.0137 \\
    RNN       & 2.2196$\pm$0.5320 & 751516  & 0.4571$\pm$0.0166 \\
    1D-ResCNN & 1.0949$\pm$0.5293 & 313735  & 0.5934$\pm$0.0184 \\
    SCNN      & 1.2127$\pm$0.5285 & 16038324 & 0.6087$\pm$0.0187 \\
    CLEEGN    & 1.6348$\pm$0.8143 & \textbf{14043} & \textbf{0.6182$\pm$0.0188} \\
    \hline
\end{tabular}}
\end{table}

Table \ref{tb:arnn_ssvep} shows the objective evaluation of each artifact removal method on the MAMEM-SSVEP-II dataset. The low decoding performance of ICLabel method shows the importance of reference method selection. From the perspective of the MSE value, 1D-ResCNN and SCNN have better averaging fitness to the reference signals than our proposed method. Despite the fact that the MSE value is not the minimal, CLEEGN outperforms each compared method in the classification accuracy and the model size. Furthermore, there is a observable improvement in the decoding performance of 1D-ResCNN, SCNN and CLEEGN from the reference method. This indicates that the cross-subject training scheme on the neural network-based methods may prevent the potential over-correction in the offline method. 

The evaluation across the two datasets suggests an overall superiority of our proposed CLEEGN model over other existing neural network-based methods in online reconstruction. These promising results indicate the usability of CLEEGN in online training/calibration-free EEG reconstruction that truly meets the need for real-world applications of EEG-based BCI.

\subsection{Visualization of Model Fitting Process}

In Figure \ref{fig:inter_vis_Adv}, we illustrate the loss curve, along with the 10-second waveform and 60-second power spectrum of the reconstructed result from the early, middle, and convergent stages on the BCI-Challenge dataset. In Figure \ref{fig:inter_vis_Adv}(a), the red line represents the monitored batch loss of the training set, while the blue line represents the monitored validation loss evaluated at the end of each training epoch. During the training process, the phasic model weights was saved at each iteration. The sample reconstructed waveform and the power spectrum on channel Fp1 at three stages with large variance are plotted in Figure \ref{fig:inter_vis_Adv}(b) and (c) respectively. In the early stage of training, the amplitude of output signals are small and there is large energy gap between the reference and reconstruction in the spectrum. Despite the evident difference in the time domain, it shows that in this stage, some primary patterns have emerged in higher energy frequency bands. As the training process progresses, the loss and validation loss decrease gradually with the oscillation becoming more pronounced in the reconstructed results. As for the spectral visualisation, the energy gap between the reconstruction and the reference diminish by degrees as the model weights are updated. Besides, the phenomenon that the dynamic with lower frequency fit faster than the higher frequency. At the convergent stage of training, we can see both the reconstructed waveform and the power spectrum are very similar to the one in reference data.

Through the series of visual observations and analysis, in the early stage of the training process, the model tends to learn the lower-frequency components, which correspond to the long-term trends in the EEG signals. In the later stage of training, there is minor changing in the gradient, and the model can learn the higher-frequency components. Under this weight updating behaviour, the loss decrease efficiently in the early stage and the fitness in both time and frequency domains improve after the training process stabilizes.

\begin{figure*}[ht]
    \centering
    \includegraphics[width=1.0\linewidth]{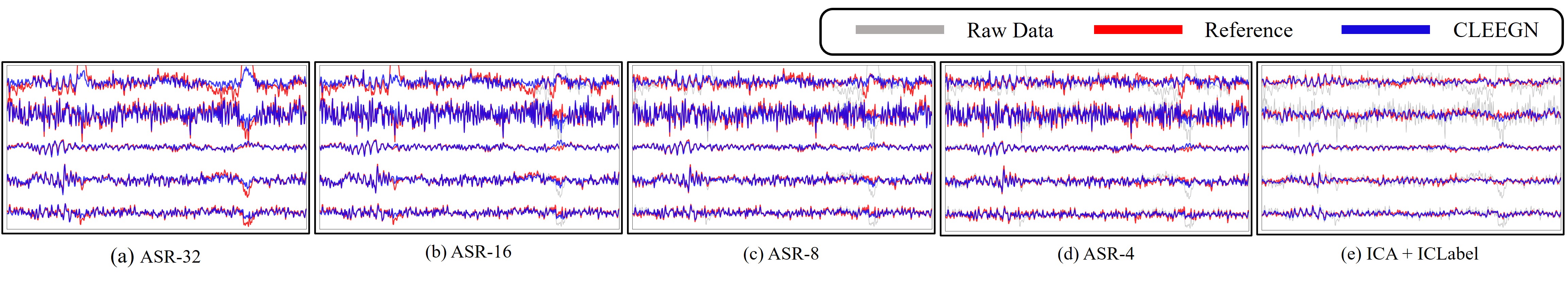}
    \caption{Visualization of raw (gray), reference (red), CLEEGN (blue) EEG waveforms with offline methods by ICLabel, ASR-32, ASR-16, ASR-8, ASR-4. Each segment plots a five-second segment of signals at Fp1, T7, Cz, T8, and O2.}
    \label{fig:cleegn_off}
\end{figure*}

\subsection{Method Tuning}

\subsubsection{Types of Reference Data}

As ICLabel, ASR can be employed to automatically generate large-scale noiseless reference EEG data offline, it is of our interest to investigate what type of reference EEG serves as the best noiseless reference data for the CLEEGN model training.

In Figure \ref{fig:cleegn_off}, subject 2's EEG waveform from the BCI-Challenge dataset is displayed. Each sub-figure shows the difference between noisy EEG data, reference EEG data, and the reconstructed EEG using the CLEEGN model. The EEG time series, representing Fp1, T7, Cz, T8, and O2 channels, highlights distinctions.
Comparing the reference waveforms reconstructed by ICA+ICLabel and ASR, ICA+ICLabel categorizes more high-frequency components as artifacts, suppressing their impact. Increasing the ASR method's cutoff parameter $k$ enhances tolerance for large-amplitude ocular artifacts, reducing differences from raw EEG data. Concerning CLEEGN's reconstructed EEG, characteristics align closely with the reference method. The ASR-trained model preserves more high-frequency components than the ICA+ICLabel-trained model. Notably, CLEEGN effectively mitigates large amplitude artifacts not identified by ASR-32 and ASR-16.

\begin{figure}[ht]
    \centering
    \includegraphics[width=1.\linewidth]{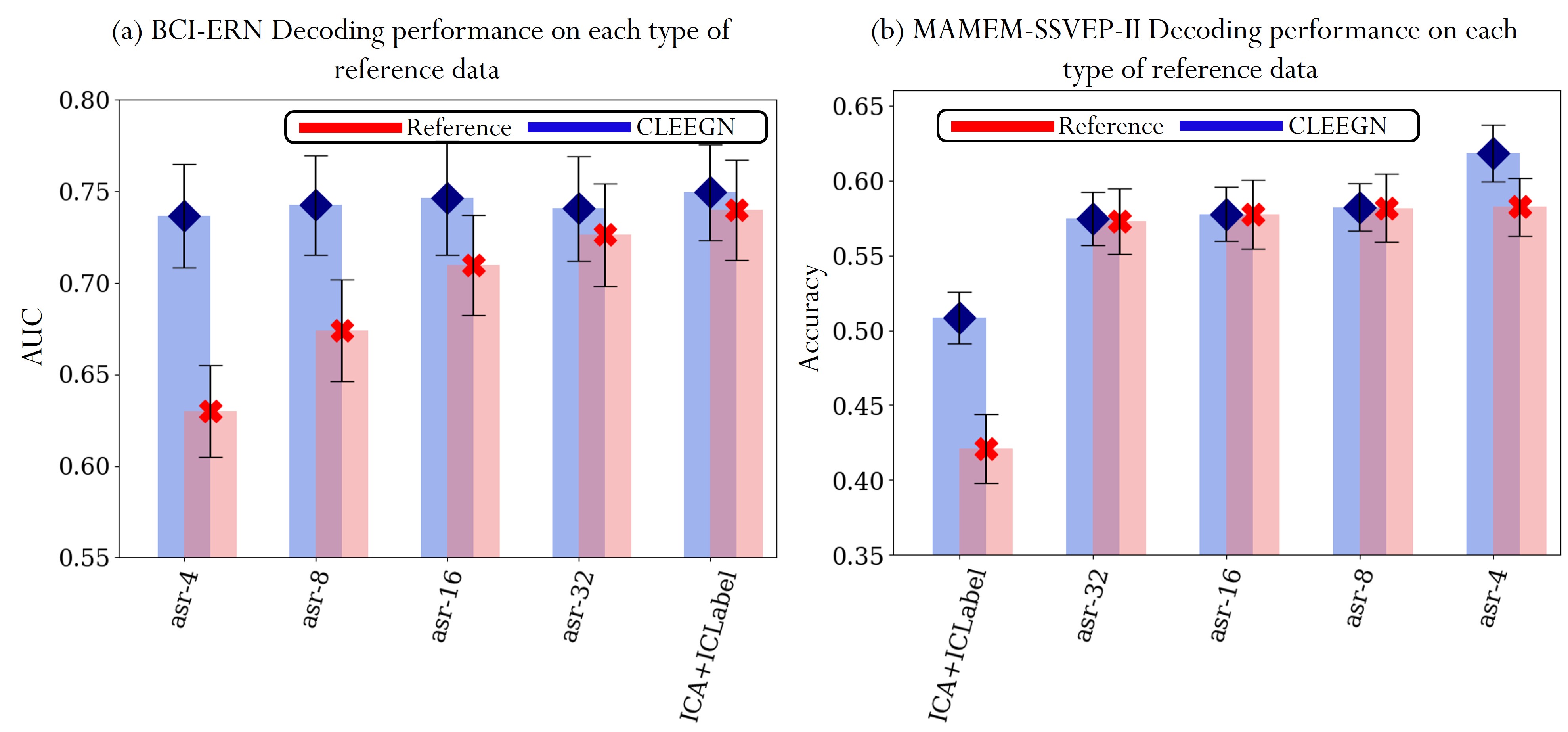}
    \caption{(a) Decoding performance of the CLEEGN-reconstructed EEG data (blue) and the corresponding reference data used for CLEEGN model training (red) using the BCI-Challenge dataset. (a) Decoding performance of the CLEEGN-reconstructed EEG data (blue) and the corresponding reference data used for CLEEGN model training (red) using the MAMEM-SSVEP-II dataset.}
    \label{fig:off_eva}
\end{figure}

Figure \ref{fig:off_eva}(a), (b) present the decoding performance between CLEEGN-reconstructed EEG data and the corresponding reference data used for CLEEGN model training using two different real-world EEG datasets. As illustrated in Figure \ref{fig:off_eva}(a), CLEEGN leverage the better performance than the reference method used in training stage on the BCI-Challenge dataset. The observation that the performance decreases with a smaller cutoff parameter $k$ in ASR method, our method shows improvement compared to the selected reference method. As for the result on MAMEM-SSVEP-II dataset, we can see the great performance gap between the ICA+ICLabel and ASR method, indicating that automatic artifact detection through ICLabel may not be a suitable on this dataset. Despite that, the reconstructed EEG from our method can improve the decoding performance from the ICA+ICLabel reference method. Though the improvement between the CLEEGN and the corresponding ASR reference method is slight, the result suggests that our method not only removes the artifact but also preserves informative brain activity in the EEG under our cross-subject training scheme.

\subsubsection{Length of Training Data per Subject}

\begin{figure}[ht]
    \centering
    \includegraphics[width=1.0\linewidth]{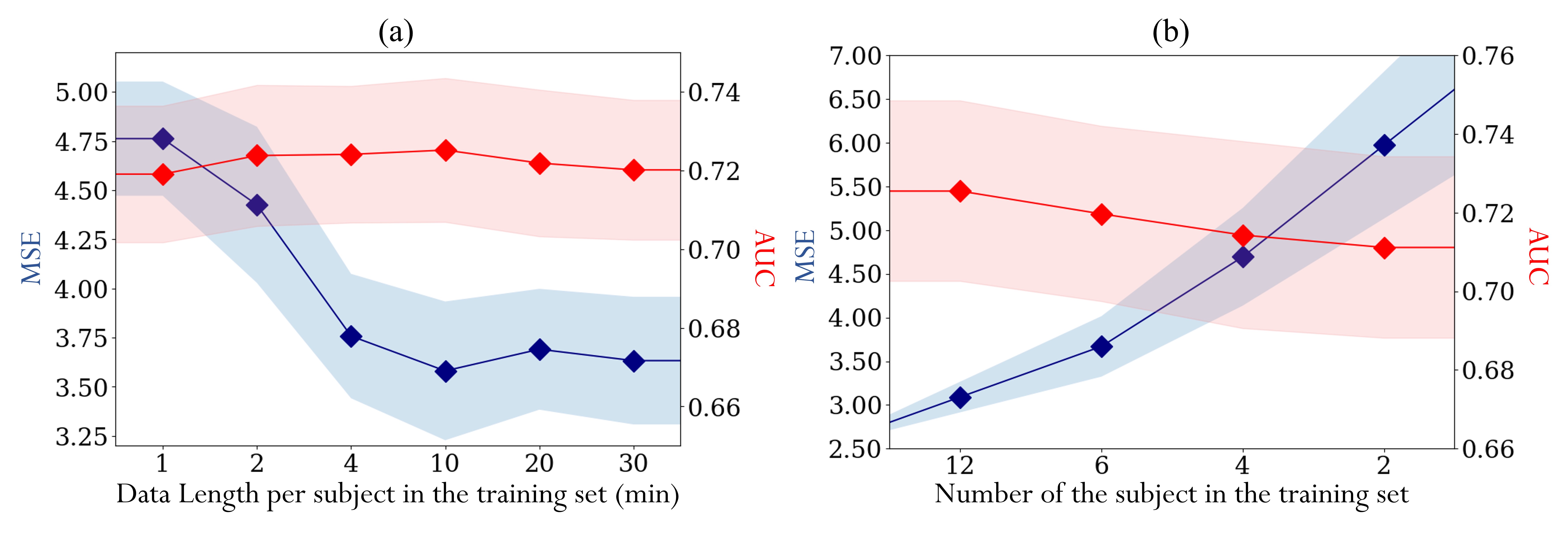}
    \caption{(a) Performance of CLEEGN against the training data length per subject evaluated by the BCI-Challenge dataset on the fitness to the reference data (blue) and the decoding performance (red). (b) Performance of CLEEGN against the number of subject in training set evaluated by the BCI-Challenge dataset on the fitness to the reference data (blue) and the decoding performance (red).}
    \label{fig:dtlen_nsbj}
\end{figure}

Subsequently, to explore the impact of training data size on reconstruction performance, the CLEEGN model undergoes training with varied durations of EEG recordings from each subject. This investigation is carried out on the BCI-Challenge dataset. For each training subject, continuous segments of 1, 2, 4, 10, 20, and 30 minutes are collected from their EEG recordings to construct the new training dataset. The blue line in Figure \ref{fig:dtlen_nsbj}(a) represents the MSE value between the reference data and the CLEEGN reconstruction data. Notably, the MSE value significantly decreases from 1 to 10 minutes, indicating improved fitness of the CLEEGN model with more data. The model trained with the initial 10 minutes achieves the minimal MSE value among all duration configurations, suggesting further minimization becomes challenging even with additional training data exceeding 10 minutes.

As for the decoding performance, the red line Figure \ref{fig:dtlen_nsbj}(b) exhibits the comparison between different training configurations. Unlike the fitness curve, the variance of BCI-Challenge classification results is relatively small. The 10-minute training setting yields the optimal decoding performance. The classification result slightly edge down while increasing or decreasing in the training data size. Interestingly, with only one-minute training data from each subject, CLEEGN can improve the decoding performance from the raw data. This indicates that CLEEGN retains its performance even when each subject only contributes a short recording for training.

\subsubsection{Number of Subject in Training Set}


In addition, we explore the effect of the number of subjects included in the training set on the fitness of CLEEGN model training and the decoding performance of the reconstructed EEG data. We randomly reduced the number of subjects for training from 12 to 6, 4, and 2. With the decrease in the number of subjects, the MSE value (blue line) and the standard error (light blue span area) increase in Figure \ref{fig:dtlen_nsbj}(b), which indicates that the generalization ability of the CLEEGN model reduces when fewer subjects are included for training. In Figure \ref{fig:dtlen_nsbj}(b), we can observe a slight decrease in decoding performance. We consider the number of subjects as an essential factor in the performance of CLEEGN.

\subsection{Feature Visualization in CLEEGN}


CLEEGN is a neural network-based model for artifact removal. In this section, we interpret the model behaviour by visualizing the 2-D mapping of the activation from each layer through a well known dimensionality reduction method, t-distributed Stochastic Neighbor Embedding (t-SNE) \cite{tsne2008desc}. t-SNE is a useful tool for generating the low-dimensional embedding from the high-dimensional data by minimizing the Kullback-Leibler divergence between the joint probabilities before and after the transformation.

\begin{figure}[ht]
    \centering
    \includegraphics[width=1.0\linewidth]{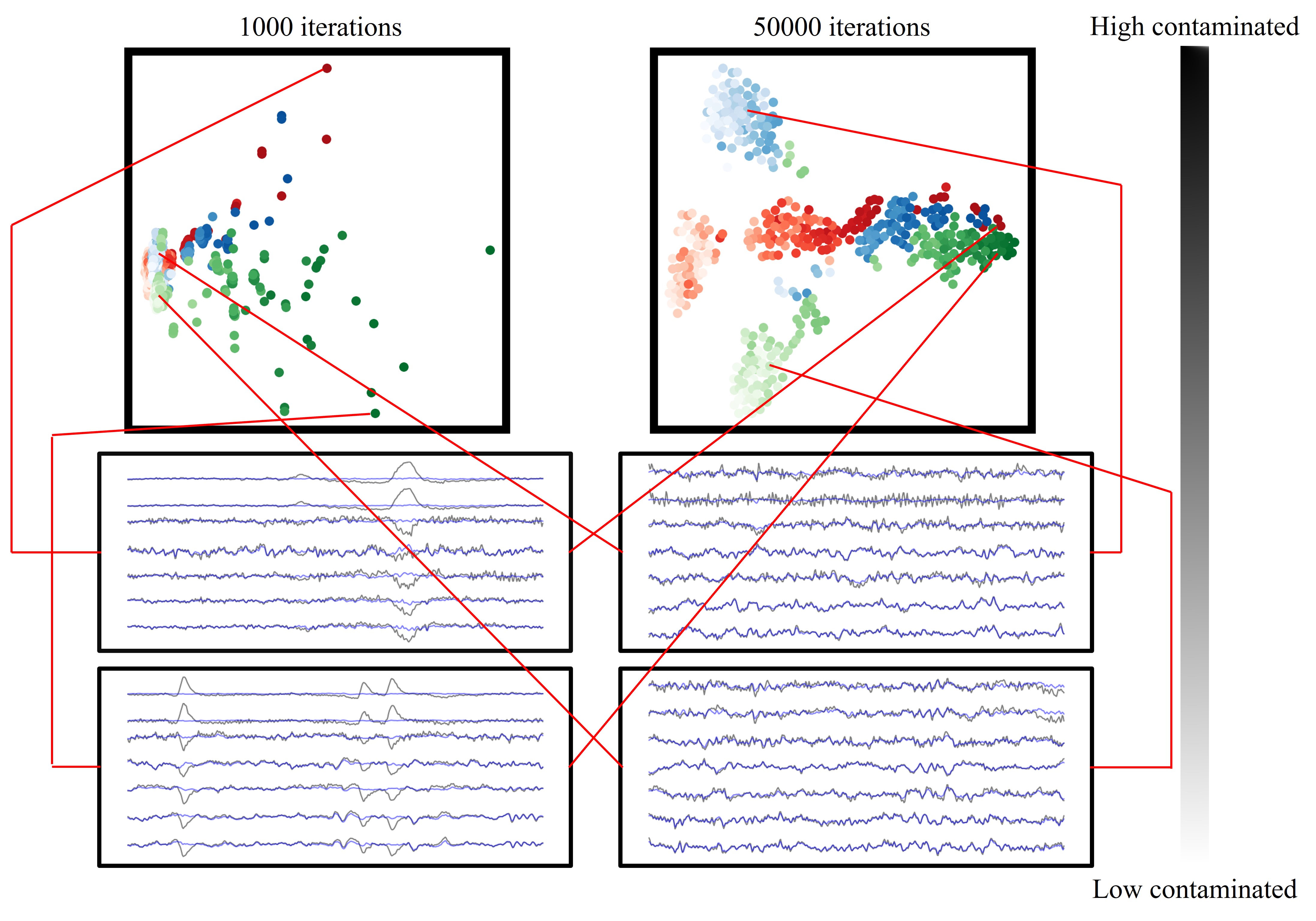}
    \caption{t-SNE visualization on the covariance matrix of multiple EEG trials.}
    \label{fig:tsne_x0_w_sample}
\end{figure}

Figure \ref{fig:tsne_x0_w_sample} shows the t-SNE visualization on the covariance matrix of a collection of EEG trials from three different subjects in BCI-Challenge dataset before processed by model. Instead of applying t-SNE to the EEG trials directly, we utilize the inter-channel covariance matrix to capture the dynamic between the brain regions in the EEG signals \cite{Cls_of_CM_4_BCI, Sleep_EEG_anz_on_CM}. Two sub-figure present the early and the steady visualized result respectively. We color the trials of three subject into red, blue and green, and the color intensity from dark to light represents the noise level based on the calculation of MSE. In the visualized result of early iteration, we can observe the EEG trails with less noise form a primary cluster, and the remaining trials generate multiple clusters in the graph along with the trials of similar contaminated level. As for the convergent state of t-SNE, three clusters consist of noiseless EEG trials in each of the three subject, while highly polluted trials from all subjects form the fourth cluster. In the both stage, we can also observe that trials of same subject tend to be closer within each cluster in the low-dimensional embedding.

\begin{figure}[ht]
    \centering
    \includegraphics[width=1.0\linewidth]{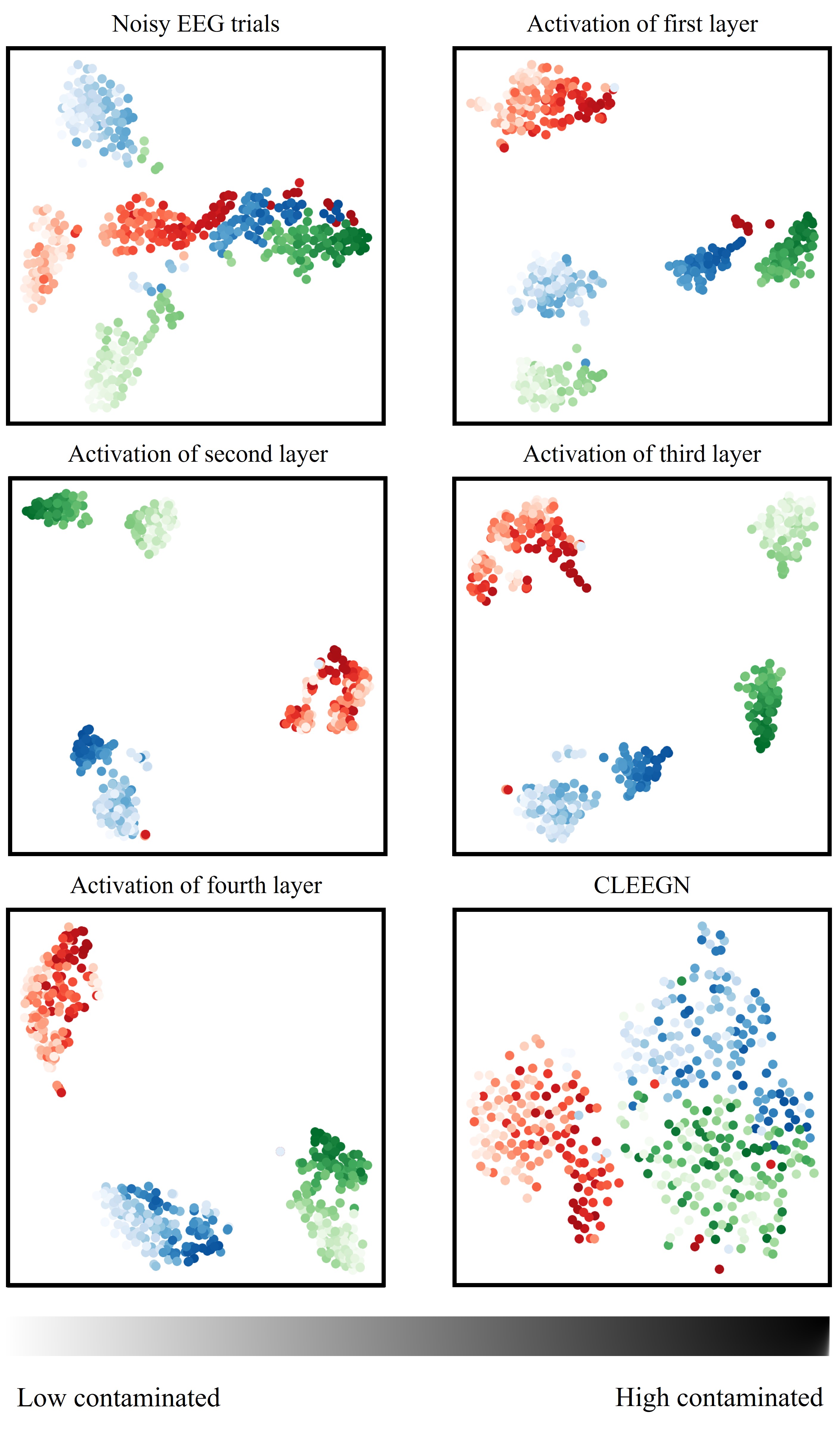}
    \caption{Visualization of the activation of the input at each layer through the t-SNE method. Each point is a 4 seconds EEG epoch. The red, green and blue represent for different subjects. The dark to light represents the noise level from high to low based on the calculation of MSE.}
    \label{fig:tsne_x012345}
\end{figure}

Figure \ref{fig:tsne_x012345} provides t-SNE visualizations of noisy EEG trials and features generated by each convolutional layer in the CLEEGN architecture, including the reconstructed results. In the first layer's latent features, noisy and noiseless trials of two subjects (colored in blue and green) form four distinct clusters. Notably, many red-colored noisy trials align with noiseless trials within the same cluster. In the visualizations of the second and third layer's latent features, green-colored trials still exhibit two clear clusters based on noise level, with a gradual merging of darker and lighter colored clusters of red and blue trials. However, distinguishing between noisy and noiseless trials becomes challenging at this stage for these two subjects. The fourth convolutional layer markedly differs from previous layers in the visualized results. While trials from different subjects still form separate clusters, each cluster contains samples with both high and low contamination, indicating significant noise suppression. Compared to the visualized latent features of each hidden layer, the outcomes of our method and noisy trials show relatively less distinct clusters in the t-SNE low-dimensional embedding. In the visualizations of reconstructed trials, subjects in different colors cluster separately, and the distinction between noisy and noiseless trials becomes indiscernible. These t-SNE visualizations suggest that CLEEGN progressively removes artifacts and reconstructs clean EEG signals layer by layer.

\section{Conclusion}

In this study, we introduce CLEEGN, a novel convolutional neural network designed for plug-and-play automatic EEG reconstruction. Our network architecture considers both spatial and temporal characteristics of EEG data. CLEEGN's performance is validated on synthetic multivariate data and two real-world EEG datasets. On the synthetic dataset, through waveform analysis and multiple objective estimations, our model effectively reconstructs features in both time and frequency domains. For real-world EEG datasets, CLEEGN training utilizes conventional offline denoising methods (ASR and ICLabel) with automatic component classification to generate abundant noiseless EEG data, employing a subject-independent scheme. The visualized waveforms demonstrate CLEEGN's ability to remove artifacts from various sources, maintaining high correlation with reference data.

Decoding performance reveals CLEEGN's superiority over other neural network-based denoising methods in ERN and SSVEP EEG reconstruction. The results suggest that CLEEGN, even without calibration, predominantly preserves inherent brain activity. Latent feature visualizations of multiple EEG trials from different subjects demonstrate model behavior and provide insights aligned with existing neuroscience knowledge. CLEEGN effectively learns the transformation of EEG reconstruction from existing techniques and surpasses conventional offline approaches. Future extensions may involve integrating other EEG denoising methods, enhancing inference capabilities across datasets or recording montages. We anticipate widespread applications of CLEEGN in future works of EEG decoding and analysis.

\bibliographystyle{IEEEtran}
\bibliography{bibtex/bib/IEEEabrv,bibtex/bib/IEEEreflist}

\appendices

\section{Neural Network Training Configuration}

\subsection{CLEEGN}

For training CLEEGN, we used the Adam optimizer with an initial learning rate of 1e-3 without weight decay. Besides, the exponential learning rate scheduler is applied with a gamma of 0.8. As for the loss function, we used Mean Squared Error (MSE). The batch size is set to 64 and the total training epoch is 40. During the training procedure, the model is evaluated using the validation subset at the end of every epoch with the purpose of saving the weights that achieved the lowest validation loss.

\subsection{IC-U-Net}

The optimizer adopted in IC-U-Net \cite{CHUANG2022119586} training is SGD with an initial learning rate of 1e-2, momentum of 0.9, and weight decay of 5e-4. The learning rate scheduler used in the training procedure is the multistep scheduler. As for the loss function, a novel ensemble loss proposed in IC-U-Net is adopted. This ensemble is a simple linear combination of the Mean Squared Error (MSE) in amplitude, velocity, acceleration, and frequency components of EEG signals. Each term in the loss function has same weight. The batch size is set to 64 and the total training epoch is 200. The weight saving strategy in IC-U-Net training is the same as CLEEGN.

\subsection{1D-ResCNN}

The Adam optimizer with an initial learning rate of 1e-3 and the Mean Squared Error (MSE) loss function is adopted in 1D-ResCNN \cite{SUN2020108} training. Since the 1D-ResCNN is developed under one-dimensional synthesized EEG data and no explicit instruction of using multi-channel EEG provided, we trained 1D-ResCNN using two different method. For one method, we trained multiple models for each EEG channel respectively. For the other method, we viewed each multi-channel EEG segment as a batch and the channel arrangement within a batch is fixed. Under our experiment, the result showed that the second method not only provides an efficient training process, but also results in a better reconstructed performance. The same weight saving strategy in CLEEGN is adopted in 1D-ResCNN training.

\subsection{EEGdenoiseNet (SCNN, RNN)}

Two further simple network architecture are adopted in the comparison, the simple convolutional structure and recurrent network structure. Mean Squared Error (MSE) loss function is adopted as the objective criterion. The weight saving strategy in SCNN and RNN training is the same as CLEEGN.

\subsection{Decoding Performance Evaluation Model: EEGNet}

EEGNet \cite{lawhern2018eegnet} is a famous EEG decoding model and widely used in EEG literature. In the original EEGNet paper, they investigated their proposed model with a different number of kernels and denoted the model with $F_1$ temporal filters and $D$ spatial filters as EEGNet-$F_1$,$D$. We use two different settings to train the two datasets used. In ERN classification, we use the EEGNet-8,2 structure suggested by the EEGNet paper. As for the SSVEP classification, an experimental result showed that EEGNet-100,8 can draw the best performance. We trained and evaluated the decoding performance individually for each subject. We divided the collection of event epochs into three splits within each subject: training set, validation set, and test set with a ratio of 4:3:3. The ratio between classes remained the same in each set. The loss function is categorical cross entropy (CCE) and the Adam optimizer is adopted with a learning rate of $10^{-3}$ and zero weight decay. The batch size is set to 32 and the total training epoch is 200.


\end{document}